\documentclass[prd,nofootinbib,superscriptaddress,preprint]{revtex4}
\usepackage[T1]{fontenc}
\usepackage{amsmath,amssymb}
\usepackage{epsfig}
\usepackage{graphicx,array}
\usepackage[usenames,dvipsnames]{color}
\usepackage{slashed}
\usepackage{comment}
\usepackage[colorlinks,citecolor=blue]{hyperref}
\usepackage{pdfpages}
\usepackage{color}
\usepackage{subfigure}
\begin{document}
\title{Neutrino mass genesis in Scoto-Inverse Seesaw with Modular $A_4$}
\author{Gourab Pathak}
\email{gourabpathak7@gmail.com}
\affiliation{Department of Physics, Tezpur University, Napaam, Assam, India-784028}
\author{Pritam Das}
\email{prtmdas9@gmail.com}
\affiliation{Department of Physics, Salbari College, Baksa, Assam 781318, India.}
\author{Mrinal Kumar Das}
\email{mkdas@tezu.ernet.in}

\affiliation{Department of Physics, Tezpur University, Napaam, Assam, India-784028}

\begin{abstract}
    
%\abstract{
We propose a hybrid scotogenic inverse seesaw framework in which the Majorana mass term is generated at the one-loop level through the inclusion of a singlet fermion. This singlet Majorana fermion also serves as a viable thermal relic dark matter candidate due to its limited interactions with other fields. To construct the model, we adopt an 
$A_4$ flavour symmetry in a modular framework, where the odd modular weight of the fields ensures their stability, and the specific modular weights of the couplings yield distinctive modular forms, leading to various phenomenological consequences. The explicit flavour structure of the mass matrices produces characteristic correlation patterns among the parameters. Furthermore, we examine several testable implications of the model, including neutrinoless double beta decay ($0\nu\beta\beta$), charged lepton flavour violation (cLFV), and direct detection prospects for the dark matter candidate. These features make our model highly testable in upcoming experiments. 
\end{abstract}
%\end{titlepage}

\maketitle 

%%%%%%%%%%%%%%%%%%% Intro %%%%%%%%%%%%%%%%%%%%
\section{Introduction}\label{sec:intro}
Neutrinos and dark matter (DM) are among the most popular and most studied yet elusive particles. Although neutrinos are placed in the standard model (SM) particle content as massless candidates, neutrino oscillation data reveals that neutrinos possess tiny masses and mixing both theoretically as well as experimentally \cite{Giunti:2003qt,ParticleDataGroup:2018ovx,Mohapatra:2005wg}. Similarly, the existence of DM in our universe is established from cosmological \cite{Planck:2018vyg,bennett2013nine} and astrophysical observations \cite{Zwicky:1937zza,rubin1970rotation}, even with the prediction of approximately $26.8\%$ of our current Universe is made up of DM. A particle being the DM candidate must be stable and electrically neutral, satisfying the current observed relic density within $0.1126 \le \Omega h^2\le 0.1246$ \cite{Planck:2018vyg} at 3$\sigma$ C.L. The SM does not allow any viable DM candidate within itself.
Therefore, we must adapt beyond the SM frameworks (BSM) to address these inconsistencies. There are numerous approaches to address light neutrino masses, and {\it seesaw} frameworks have gotten a lot of attention in past years \cite{Das:2017ski,King:2015sfk,Barrie:2022cub,Mohapatra:2004zh} over others. Many seesaw frameworks require heavy fields to address tiny neutrino masses, and the search for those heavy fields are yet another challenge for the experimentalists. Therefore, low-scale seesaw mechanisms like {\it inverse seesaw framework} \cite{Abada:2021yot,Dias:2012xp,Gautam:2020wsd,Mukherjee:2017pzq} with TeV scale additional fields can generate tiny neutrino masses, which can also be detected in colliders and a useful tool to explore extensively.  In the study of neutrino masses, two possible mass ordering patterns arise due to the current uncertainty in the sign of one of the squared mass differences. Typically, it is essential to examine both normal and inverted orderings to thoroughly assess the viability of any theoretical model. However, inspired by recent studies from various authors \cite{Heavens:2018adv, Jimenez:2022dkn, Gariazzo:2022ahe}, we choose to focus exclusively on the normal mass ordering in this work.

Parallel to the neutrino mass puzzle, the more appealing frameworks are those where both neutrino mass and dark matter are addressed under the same roof. Radiative seesaws are popular for generating neutrino mass at the loop level and addressing DM simultaneously. A hybrid framework combining tree-level and radiative seesaw has gained attention in recent years \cite{Fraser:2014yha,Mandal:2021yph}. Keeping an eye on the tree-level seesaw scale, where neutrino masses can be generated at energies at the reach of current or upcoming experimental searches ($i.e., $ around the TeVs) to test the feasibility of the model, we adopt the inverse seesaw framework with radiative correction \cite{Mandal:2019oth}. In this framework, the Majorana mass term is restricted at the tree level via $ad-hoc$ global $U(1)_{B-L}$ charge and the Majorana mass term $``\mu"$ is generated at the loop-level to complete the inverse framework. To generate the Majorana mass term in one loop, we have introduced another singlet fermion, which plays the role of dark matter in our analysis. The essence of this framework lies in the fact that it simplifies the calculations involved in the radiative seesaw framework, and the DM candidate itself embryos the genesis of neutrino mass and mixing.  The study of neutrinoless double beta decay ($0\nu\beta\beta$) offers a unique opportunity to confirm the Majorana nature of neutrinos. If established, it can provide definitive insights into the absolute neutrino mass scale. In $0\nu\beta\beta$, the lepton number is significantly violated, resulting in the emission of a pair of electrons. On the other hand, the charged lepton flavour violation (cLFV) refers to processes where a charged lepton transitions into another charged lepton of a different flavour, violating lepton flavour conservation.

To execute and complete the framework, we have adopted the well-studied modular symmetry approach \cite{Feruglio:2017spp,King:2020qaj,Kobayashi:2018vbk}, where the modulus $\tau$ acquires a non-zero VEV by breaking the flavour symmetry. Introducing complex modulus $\tau$ minimizes the use of additional flavon fields and keeps a rigid grip on the Yukawa couplings, which have a definite form depending on the modular weight $k_I$. The modular forms of the Yukawa couplings transform non-trivially and are holomorphic functions of $\tau$. An extensive study has been carried out with modular symmetry considering discrete flavour symmetries such as $S_3$ \cite{Okada:2019xqk,Meloni:2023aru}, $S_4$ \cite{Kobayashi:2019xvz,Wang:2019ovr,Penedo:2018nmg,Zhang:2021olk},$A_4$ \cite{Kashav:2021zir,Kashav:2022kpk,Singh:2024imk,Mishra:2023cjc,Nomura:2019jxj,Nomura:2019xsb,Gogoi:2023jzl,Behera:2020lpd,Abbas:2020qzc,Altarelli:2005yx,Das:2018qyt}, $A_5$ \cite{Novichkov:2018nkm,Ding:2019xna,Behera:2021eut,Behera:2022wco}. There are some popular works with $A_4$ in the extended seesaw frameworks \cite{Kumar:2024zfb,Kumar:2023moh,Behera:2020sfe,CentellesChulia:2023osj,Mishra:2022egy}, which motivate the adoption of modular $A_4$ flavour symmetry in this work. The odd modular weight of a particle essentially demands odd coupling (either Yukawa or quartic), which then restricts the interaction. This behaviour is quite identical with the discrete $Z_2$ odd charge. Hence, the one with odd modular weight(s) is inherently stable without the requirements of any additional $Z_n$ symmetries.

In this work, we adopted the non-susy approach of modular symmetry\footnote{The holomorphic nature of modular symmetry favours the supersymmetric approach, yet we are working on the non-SUSY framework by writing the Lagrangian rather than superpotential \cite{Nomura:2019jxj, Nomura:2019xsb}. In our case, we have three fields $N_R$, $f$ and $\eta$ with non-zero modular weights, and one can write similar types of terms in superpotential format without altering the phenomenology of the current model.} using the $A_4$ flavour symmetry to construct our model. 
The inverse seesaw framework is adopted to generate neutrino mass, with the exception of generating the Majorana mass term via the radiative correction. The odd modular weight stabilizes the dark matter candidate, which is also associated with the genesis of the Majorana mass term at 1-loop level. We study the scenario of fermionic dark matter by calculating the thermal relic. Various detection prospects, including the neutrinoless double beta decay ($0 \nu \beta \beta$), charged lepton flavour violation (cLFV) for $\mu\rightarrow e\gamma$, $\tau\rightarrow e\gamma$ along with the direct detection in DM searches, are also checked to establish the concreteness of framework. The novelty of this work lies in the implementation of modular flavour symmetry within the scoto-inverse seesaw framework, a combination that, to the best of our knowledge, has not been explored before. Previous studies have either applied modular flavour symmetry to the scotogenic framework \cite{Behera:2020lpd}, utilized the scoto-inverse seesaw mechanism with discrete flavour symmetries \cite{Mandal:2019oth}, or embedded the inverse seesaw framework in modular flavour symmetry settings \cite{Nomura:2019xsb}. We implement the scoto-inverse seesaw framework, where the Majorana mass term is generated radiatively at the one-loop level through newly introduced fermion singlet. This singlet, possessing odd modular weights, naturally stabilizes and serves as a dark matter (DM) candidate. As a result, the stability of the DM candidate is ensured without the need for any additional discrete symmetry.

Our paper is organized as follows. In section \ref{sec2}, we explicitly addressed all the components of the model framework. Numerical analysis for both the neutrino sector and the dark matter sector is discussed in section \ref{sec3}. Detection possibilities of various processes or particles involved in this study is carried out in section \ref{sec4} and finally we concluded our study in section \ref{sec5}. 

%%%%%%%%%%%%%%%%%%%%% Model Framework %%%%%%%%%%%%%%%%%%%%%%%%%
%%%%%%%%%%%%%%%%%%%%%%%%%%%%%%%%%%%%%%%%%%%%%%%%%%%%%%%%%%%%%%%

\section{MODEL FRAMEWORK}
\label{sec2}
In this section, we provide an overview of the model framework for the scotogenic inverse seesaw mechanism. The particle content and group charges are provided in Table \ref{tab1}. We prefer to extend using discrete $A_{4}$ modular symmetry to investigate neutrino phenomenology, and we impose a global $U(1)_{B-L}$ symmetry to avoid certain undesirable interactions. The SM particle spectrum is extended with three right-handed ($N_R$) and three left-handed ($S_L$) heavy fermions. We propose that the dark sector serves as the origin of neutrino mass. To achieve this, we have introduced a dark fermion $f$ along with a complex dark scalar $\eta$. The assigned modular weight effectively emulates $Z_2$ symmetry, preventing neutrino mass at the tree level and stabilizing dark matter. Additionally, Table \ref{tab1} provides the non-trivial transformations of Yukawa couplings, scalar couplings, and their corresponding modular weights. In the following subsections we briefly introduced the scalar and the fermionic sector and the terms associated with it. 

\begin{table}[h]\footnotesize
    \centering
    \begin{tabular}{|c|ccccccc||cccccc|}
    \hline
Symmetry&&&Particle content&&&&&&&Couplings&&&\\
   % \hline
     Charges      & $L_{l}$ & $e_R$ & $N_R$ & $S_L$ & $f$ & $H$ & $\eta$ & Y  & $Y_{3a}^{(6)}$ & $Y_{1}^{(6)}$ & $\rho$ & $\lambda_{\eta}$ & $\lambda_{\eta}^{'}$\\
         \hline
         $SU(2)_L$ & 2  & 1 & 1 & 1 & 1 & 2 & 1 &-&-&-&-&-&-\\
         $U(1)_Y$& -1 & -2 & 0 & 0 & 0 & 1 & 0 &-&-&-&-&-&-\\
         $U(1)_{B-L}$& -1 & 1 & 1 & -1 & 0 & 0 &1 &-&-&-&-&-&- \\
         $A_4$ & $(1,1'',1')$ & $(1,1',1'')$ & 3 & 3 & 1 & 1 &1 & 3  & 3 & 1 & 3 & 1 & 1 \\
         $k_{I}$& 0 & 0 & -2 & 0 & -3 & 0 & -3 & 2  & 6 & 6 & 2 & 12 & 6 \\
    \hline
    \end{tabular}
    \caption{Particle content of the model and the couplings with their $A_4$ and modular charges $k_I$.}
    \label{tab1}
\end{table}

\subsection{The scalar}
The scalar potential of the model is given as:\\
\begin{equation}\label{pot1}
    \mathcal{V}=-m_{H}^2(H^{\dag}H)+\frac{\lambda_{H}}{2}(H^{\dag}H)^2+\frac{\lambda_{\eta}}{2}\zeta_{1}(\eta^*\eta)^2+\lambda_{\eta}^{'}[-m_{\eta}^2(\eta^*\eta)+\zeta_{2}(H^{\dag}H)(\eta^*\eta)+\frac{\mu_{\eta}^2}{4}(\eta^2+(\eta^{*})^{2})].
\end{equation}
Here, $H=\begin{pmatrix}
    G^+ \\
    \frac{v+h+iA}{\sqrt{2}}
\end{pmatrix}$ is the SM Higgs doublet with $v$ being the VEV and $\eta = (\eta_{R} + i \eta_{I})/\sqrt{2}$ is the complex dark scalar. In the above potential $\zeta_{i}$'s are the free parameters and the scalar coupling $\lambda_{\eta}$ and $\lambda_{\eta}^{'}$ are the singlet representation of $A_4$ with modular weight 12 and 6 respectively, which can be expressed in terms of the components of weight-2 triplet Yukawa couplings (see Eq. \eqref{eq:weight12} and Eq. \eqref{eq:weight6}):
\begin{eqnarray}
\nonumber\lambda_{\eta}^{'}&&=y_{1}^3+y_{2}^3+y_{3}^3-3y_{1}y_{2}y_{3},\\
    \lambda_{\eta}&&=(y_{1}^2+2y_{2}y_{3})^3.
\end{eqnarray}
To ensure the stability of the scalar potential from below, we adhere to the following conditions of the quartic couplings at any given energy scale $``\Lambda"$:
\begin{eqnarray}
\lambda_H(\Lambda)>0;~~\lambda_\eta(\Lambda)>0;~~\lambda_\eta^\prime (\Lambda)+\sqrt{\lambda_H(\Lambda)\lambda_\eta(\Lambda)}>0.
\end{eqnarray}
We also keep a strict eye on the perturbativity limit of the quartic couplings associated with the scalar potential \eqref{pot1}, $i.e.,$ $\lambda_H,\lambda_\eta,\lambda_\eta^\prime\le 4\pi$.

The mass spectrum of the scalar sector is given by:
\begin{equation}
\begin{aligned}
m_{h}^2 &= \lambda_{H} v^{2},\\
m_{\eta_{R}} &= \lambda_{\eta}^{'}[m_{\eta}^2 +\frac{1}{2}
(\zeta_{2} v^{2}+\mu_{\eta}^2)],\\
m_{\eta_{I}} &= \lambda_{\eta}^{'}[m_{\eta}^2 +\frac{1}{2}
(\zeta_{2} v^{2}-\mu_{\eta}^2)].
\end{aligned}
\end{equation}

\subsection{The fermion part: Scotogenic Inverse 
seesaw mechanism for neutrino mass generation}
We adopt the scotogenic inverse seesaw mechanism
for neutrino mass generation. 
Within the current model, which incorporates \(A_4\) modular symmetry, the complete \(9 \times 9\) neutral fermion mass matrix for the inverse seesaw mechanism in the flavor basis \((\nu_{L}, N_{R}, S^c_{L})\) is given by
%\begin{equation}
 %   \mathbb{M} =  \begin{array}{c|ccc}
  %       & \nu_L & N_{R}^c & S_L \\
   %      \hline
   % \nu_L & 0 & M_D & 0\\
    %N_{R}^c & M_{D}^T & 0 & M_{R}^T\\
    %S_L & 0 & M_R & \mu
    %\end{array}
%\end{equation}
\begin{eqnarray}
   \mathbb{M}=  \begin{pmatrix}
        0&M_D&0\\M_D^T&0&M_R^T\\0&M_R&\mu
    \end{pmatrix}.\label{eq:M matrix}
\end{eqnarray}
Using the appropriate mass hierarchy among the mass matrices as provided below:
\begin{equation}
    M_{R}>>M_{D}, \, \mu.
\end{equation}
The light neutrino mass in inverse seesaw is given by:
\begin{equation}
    m_{\nu} = M_D{M_R}^{-1}\mu{M_R^{-1}}^TM_D^T.
    \label{eq:nu mass}
\end{equation}

 Here, $M_D$ is the $3\times 3$ Dirac mass term and $M_R$ is $3\times 3$ the mixing terms for the $N_R-S_L$ mixing. The $\mu$ term is missing at tree level and it is generated at one loop-level via radiative correction. In the following subsections, we briefly introduce the respective interaction terms with their mass matrices.    
\begin{figure}
    \centering
    \includegraphics[scale=0.71]{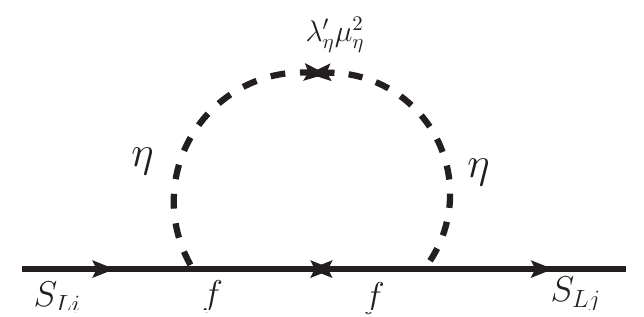}
    \caption{One loop contribution to neutrino mass via scotogenic inverse seesaw}
    \label{fig:fig-1}
\end{figure}
\subsubsection{Dirac mass term for charged leptons ($M_{l}$):}
To achieve a simplified structure for the charged lepton mass matrix, we consider the left-handed lepton doublets \{$L_{e_{L}}, L_{\mu_{L}}, L_{\tau_{L}}$\} transforming under the $A_{4}$ group as \{$1, 1^{'}, 1^{''}$\} and the right-handed charged leptons as \{$1, 1^{''}, 1^{'}$\}. The Standard Model Higgs doublet $H$ is considered a singlet under the $A_4$ group. All these fields are assigned zero modular weight. Under $U(1)_{B-L}$ the left-handed lepton doublets, right-handed charged leptons, and the Higgs doublet have charges -1, 1, and 0, respectively. The relevant interaction Lagrangian term for charged leptons is given by:
\begin{equation}
    \mathcal{L}_{M_{l}} = y_{l}^{ee}\bar{L}_{eL}He_{R} + y_{l}^{\mu\mu}\bar{L}_{\mu L}H\mu_{R} + y_{l}^{\tau\tau}\bar{L}_{\tau L}H\tau_{R}.
\end{equation}
After spontaneous symmetry breaking, the charged lepton mass matrix becomes:
\begin{equation}
    M_{l} = \begin{pmatrix}
       \frac{y_{l}^{ee} v}{\sqrt{2}} & 0 & 0 \\
       0 & \frac{y_{l}^{\mu\mu} v}{\sqrt{2}} & 0 \\
       0 & 0 & \frac{y_{l}^{\tau\tau} v}{\sqrt{2}} \\
       \end{pmatrix}
       =\begin{pmatrix}
           m_{e} & 0 & 0\\
           0 & m_{\mu} & 0\\
           0 & 0 & m_{\tau}
       \end{pmatrix}.
\end{equation}

\subsubsection{Dirac neutrino mass term connecting $\nu_{L}$ and $N_{R}$ ($M_{D}$):}
We consider the left-handed lepton doublets \{$L_{e_{L}}, L_{\mu_{L}}, L_{\tau_{L}}$\} transforming under $A_{4}$ as  \{$1, 1^{''}, 1^{'}$\} and right-handed neutrinos ($N_{R}$) as triplet 3. The SM Higgs doublet $H$ is transforming under $A_{4}$ as a singlet. Then the Dirac Yukawa term $\bar{L}_{L}\Tilde{H}N_{R}$ is allowed under $A_{4}$ symmetry and the corresponding Yukawa coupling transforming under $A_{4}$ modular group as triplets 3 shown in Table \ref{tab1}. For the modular forms of weight 2, $Y= (y_{1}(\tau), y_{2}(\tau), y_{3}(\tau))$ is given in terms of Dedekind eta-function $\eta(\tau)$ and its
derivative \cite{Altarelli:2005yx}, where $\tau$ is a complex number. Hence, the invariant Dirac Lagrangian can be represented in the following way:
\begin{equation}
    \mathcal{L}_{M_{D}} = \alpha_D\bar{L}_{eL}\Tilde{H}(YN_R)_{1} + \beta_D\bar{L}_{\mu L}\Tilde{H}(YN_R)_{1'} + \gamma_D\bar{L}_{\tau L}\Tilde{H}(YN_R)_{1''},
\end{equation}
where, \{$\alpha_{D}, \beta_{D}, \gamma_{D}$\} are free parameters. After spontaneous symmetry breaking, the Dirac neutrino mass matrix becomes:
\begin{equation}
    M_D = \frac{v}{\sqrt{2}} \begin{pmatrix}
          \alpha_D & 0 & 0 \\
          0 & \beta_D & 0 \\
          0 & 0 & \gamma_D\\
        \end{pmatrix}.
        \begin{pmatrix}
            y_1 & y_3 & y_2 \\
            y_2 & y_1 & y_3 \\
            y_3 & y_2 & y_1 \\
        \end{pmatrix}.
\end{equation}

\subsubsection{Mixing term connecting $N_{R}$ and $S_L$ ($M_{R}$):}
Both types of sterile neutrinos $N_{R}$ and $S_L$ transform as triplet 3 under $A_{4}$ in our study. The mixing term $\bar{S}_{L}N_{R}$ is allowed due to charge assignment of $U(1)_{B-L}$ and it is given as follows:
\begin{equation}
    \mathcal{L}_{M} = \rho[\alpha_{NS}(\bar{S}_{L}N_R)_{Symm}+\beta_{NS}(\bar{S}_{L}N_R)_{Anti-symm}].\label{eqLM}
\end{equation}
Here, $\alpha_{NS}$ and $\beta_{NS}$ are free parameter. The first term Eq. \eqref{eqLM} is symmetric and second term is anti-symmetric product for $\bar{S}_{L}N_{R}$ making $3_{s}$ and $3_{a}$ representations of $A_{4}$. $\rho$ is the coupling which is triplet 3 under $A_{4}$ and can be written in the following way:

\begin{equation}
    [\rho_{N_{1}}(\tau), \rho_{N_{2}}(\tau), \rho_{N_{3}}(\tau)]^{T}=\rho_{0}[y_{1}(\tau), y_{2}(\tau), y_{3}(\tau)]^{T}.
\end{equation}
where $\rho_{0}$ is a free parameter determining the scale of right-handed neutrino mass. The resulting mass matrix is found to be,
\begin{equation}
    M_R = \rho_{0}\frac{\alpha_{NS}}{3}\begin{pmatrix}
        2y_1 & -y_3 & -y_2 \\
        -y_3 & 2y_2 & -y_1 \\
        -y_2 & -y_1 & 2y_3 \\
    \end{pmatrix} + \rho_0\beta_{NS} \begin{pmatrix}
        0 & y_3 & -y_2\\
        -y_3 & 0 & y_1\\
        y_2 & -y_1 & 0\\
    \end{pmatrix}.
\end{equation}

\subsubsection{Majorana mass term for $S_L$ ($\mu$):}
The Majorana mass term $\bar{S^{c}_{L}}S_L$ is not allowed due to the assigned $U(1)_{B-L}$ charge. However, this term arises only due to calculable radiative correction indicated in Fig.\ref{fig:fig-1}. This results in the neutrinos being massless at tree level and the soft-breaking term in the potential $\frac{\mu_{\eta}^2}{4}(\eta^2+(\eta^{*})^{2})$ give rise to lepton number violation and generate Majorana mass term for $S_L$ in the one-loop level following way:
\begin{eqnarray}\label{lmu}
   \nonumber \mathcal{L}_{\mu} &=& \beta_{L} Y_{3a}^{(6)} \eta f S_L + Y_{1}^{(6)} \kappa_{s} f f\\ & =& \beta_{L}(y_{31}\eta f S_{L1} + y_{33}\eta f S_{L2} + y_{32}\eta f S_{L3}) + Y_{1}^{(6)} \kappa_{s} f f,
\end{eqnarray}
where $\kappa_s, \beta_{L}$ are free parameters and the detailed structure of the Yukawas such as $Y_{3a}^{(6)}$ can be found it the appendix \ref{sec: Append B}. The \(\mu\) term can be expressed in the following way: 
\begin{equation}
\mu =  F(m_{\eta_{R}}, m_{\eta_{I}}, M_{f})M_{f}h_{i}h_{i}    ,
\end{equation}
where, $M_{f} = Y_{1}^{(6)} \kappa_{s}$ and the couplings \(h_{i}\) are given as follow:
\begin{equation}
    h_{1}= \beta_{L}y_{31}, \quad h_{2}=\beta_{L}y_{33}, \quad h_{3}=\beta_{L}y_{32} .
\end{equation}
Hence, the $\mu$ term at the loop level becomes:
\begin{equation}
    \mu = \beta_{L}^2 M_{f}\begin{pmatrix}
        y_{31}^2 & 0 & 0 \\
        0 & y_{33}^2 & 0 \\
        0 & 0 & y_{32}^2 \\
    \end{pmatrix} F(m_{\eta_{R}}, m_{\eta_{I}}, M_{f}),
\end{equation}

with, \begin{equation}
   F(m_{\eta_{R}}, m_{\eta_{I}}, M_{f}) = \frac{1}{32 \pi^2}\Big[\frac{m_{\eta_{R}}^2}{M_{f}^2 - m_{\eta_{R}}^2}\ln{\frac{M_{f}^2}{m_{\eta_{R}}^2}} - \frac{m_{\eta_{I}}^2}{M_{f}^2 - m_{\eta_{I}}^2}\ln{\frac{M_{f}^2}{m_{\eta_{I}}^2}}\Big] .
\end{equation}

%%%%%%%%%% Numerical Analysis %%%%%%%%%%%%%%%%%%%%%%%%%%%%%%%

\section{Numerical analysis}\label{sec3}
\subsection{Neutrino fit data}

In numerical analysis, we use the global fit neutrino oscillation data within a $3\sigma$ interval as given in Table \ref{tab:oscillation data}.
\begin{table}[h]
    \centering
    \begin{tabular}{||c||c|c||}
    \hline
    \multicolumn{3}{|c||}{Normal ordering}\\
    \hline
           & bfp $\pm 1\sigma$ & $ 3\sigma$ range \\
         \hline
         $\sin^2{\theta_{12}}$ & $0.303_{-0.011}^{+0.012}$  & $0.270 - 0.341$ \\
         $\sin^2{\theta_{13}}$& $0.02203_{-0.00059}^{+0.00056}$ & $0.02029 - 0.02391$ \\
         $\sin^2{\theta_{23}}$& $0.572_{-0.023}^{+0.018}$ & $0.406 - 0.620$  \\
         $\frac{\Delta m_{21}^2}{10^{-5} (eV^2)}$ & $7.41_{-0.20}^{+0.21}$ & $6.82 - 8.03$  \\
         $\frac{\Delta m_{31}^2}{10^-{3} (eV^2)}$& $2.511_{-0.027}^{+0.028}$ & $2.428 - 2.597$  \\
    \hline
    \end{tabular}
    \caption{The NuFIT 5.2 (2022) results \cite{Esteban:2020cvm}.}
    \label{tab:oscillation data}
\end{table}

To numerically diagonalize the neutrino mass matrix as given in Eq.\eqref{eq:nu mass}, we use the relation \( U^{\dagger} \mathcal{M} U = \text{diag}(m_1^2, m_2^2, m_3^2) \), where \( \mathcal{M} = m_{\nu} m_{\nu}^{\dagger} \) and \( U \) is a unitary matrix. In our model, as the charged lepton matrix is diagonal, \(U=U_{PMNS}\), is the well-established PMNS unitary matrix. The PMNS matrix is parametrized by three mixing angles \((\theta_{12}, \theta_{13}, \theta_{23})\), one Dirac phase \((\delta_{CP})\) and two Majorana phases \((\alpha_{21}, \alpha_{31})\) as follows:
\begin{equation}
 U_{PMNS} = \begin{pmatrix}
     c_{12}c_{13} & s_{12}c_{13} & s_{13}e^{-i\delta_{CP}}\\
     - s_{12}c_{23} - c_{12}s_{23}s_{13}e^{i\delta_{CP}} & c_{12}c_{23} - s_{12}s_{23}s_{13}e^{i\delta_{CP}} & s_{23}c_{13}\\
     s_{12}s_{23} - c_{12}c_{23}s_{13}e^{i\delta_{CP}} & - c_{12}s_{23} - s_{12}c_{23}s_{13}e^{i\delta_{CP}} & c_{23}c_{13}
 \end{pmatrix}  
 . \begin{pmatrix}
     1 & 0 & 0\\
     0 & e^{i\frac{\alpha_{21}}{2}} & 0\\
     0 & 0 & e^{i\frac{\alpha_{31}}{2}}
 \end{pmatrix}
\end{equation}
where, \(c_{ij}=\cos{\theta_{ij}}\) and \(s_{ij}=\sin{\theta_{ij}}\).
From this unitary matrix \( U_{PMNS} \), the neutrino mixing angles can be determined using standard formulas:
\begin{equation}
    \sin^2{\theta_{13}}=|U_{13}|^2, \, \sin^2{\theta_{12}} = \frac{|U_{12}|^2}{1 - |U_{13}|^2}, \, \sin^2{\theta_{23}}=\frac{|U_{23}|^2}{1 - |U_{13}|^2}.
\end{equation}
We also calculate the Jarlskog invariant \(J_{CP}\) and the CP-violating phase \(\delta_{CP}\) from the elements of the PMNS matrix:
\begin{equation}
    J_{CP} = Im[U_{e1}U_{\mu2}U_{e2}^{*}U_{\mu1}^{*}] = s_{23}c_{23}s_{12}c_{12}s_{13}c_{13}^{2}\sin{\delta_{CP}}.
\end{equation}
The Majorana phases are also calculated using the following two invariants:
\begin{equation}
\begin{split}
& I_{1} = Im[U_{e1}^{*}U_{e2}] = c_{12}s_{12}c_{13}^{2}\sin{\left(\frac{\alpha_{21}}{2}\right)},\\
& I_{2} = Im[U_{e1}^{*}U_{e3}] = c_{12}s_{13}c_{13}\sin{\left(\frac{\alpha_{31}}{2}-\delta_{CP}\right)}.
\end{split}    
\end{equation}

\begin{figure}[h!]
        \centering
        \includegraphics[scale=0.45]{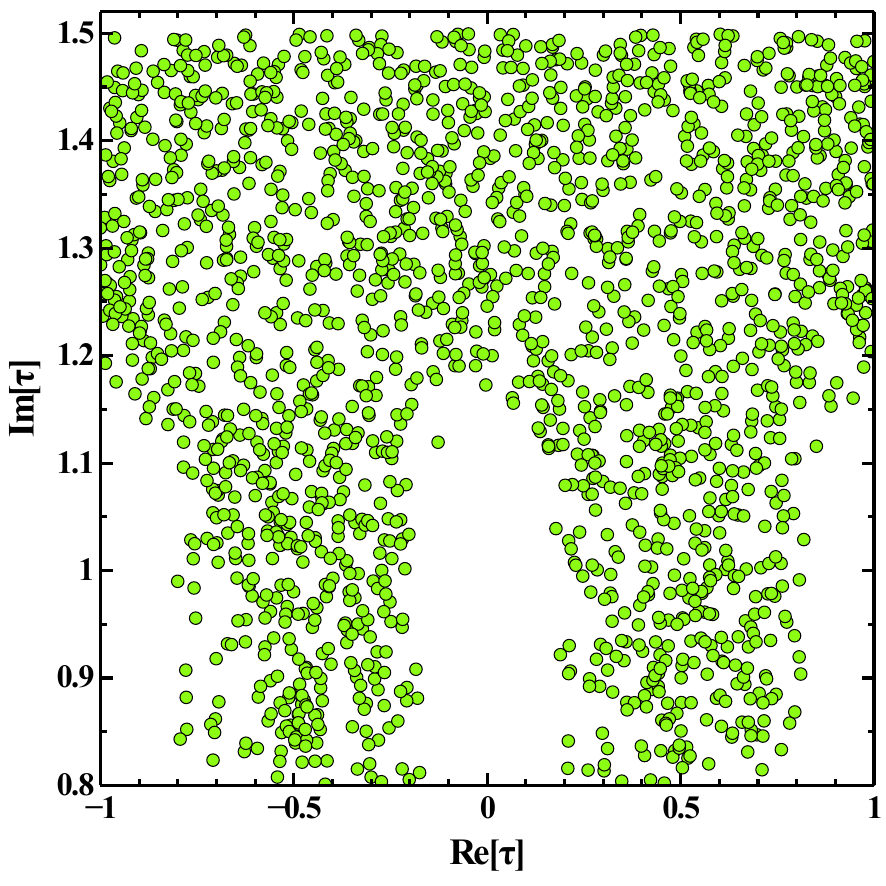}
        \includegraphics[scale=0.45]{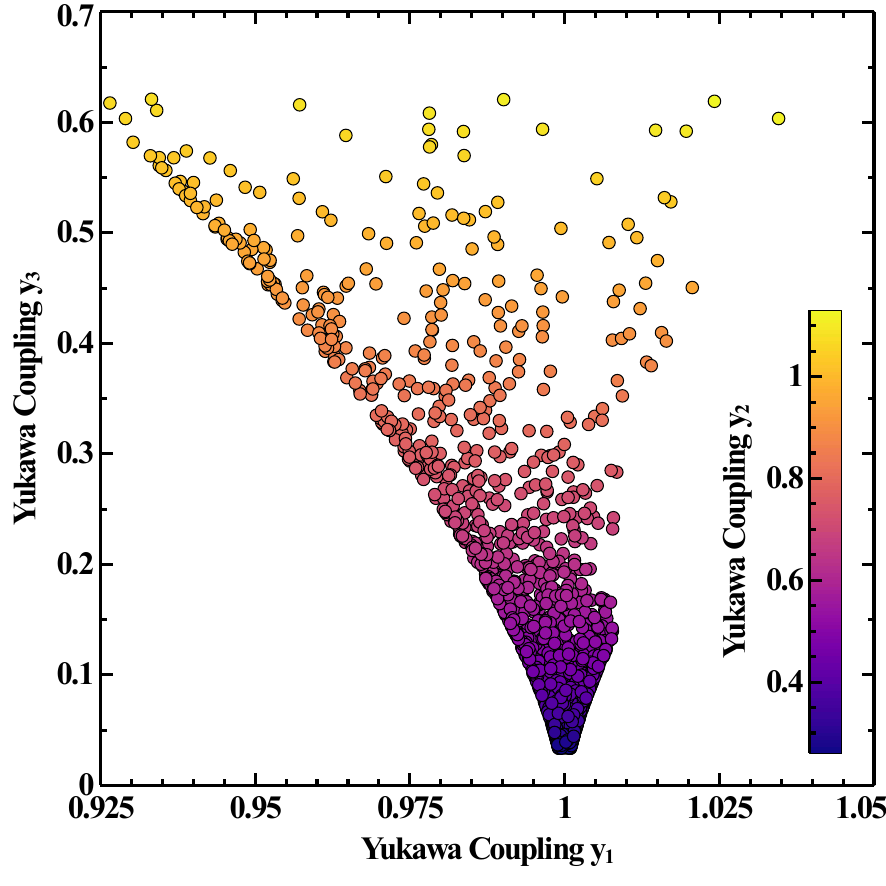}
        \caption{($Left$) Correlation between the real component and the imaginary component of $\tau$. ($Right$) Correlation among the Yukawa couplings. The colour bar projected the variation of $y_2$, while $y_1$ and $y_3$ reside in the respective axes.}
       \label{fig:fig-2} 
\end{figure}
\begin{figure}
    \centering
    \includegraphics[scale=0.45]{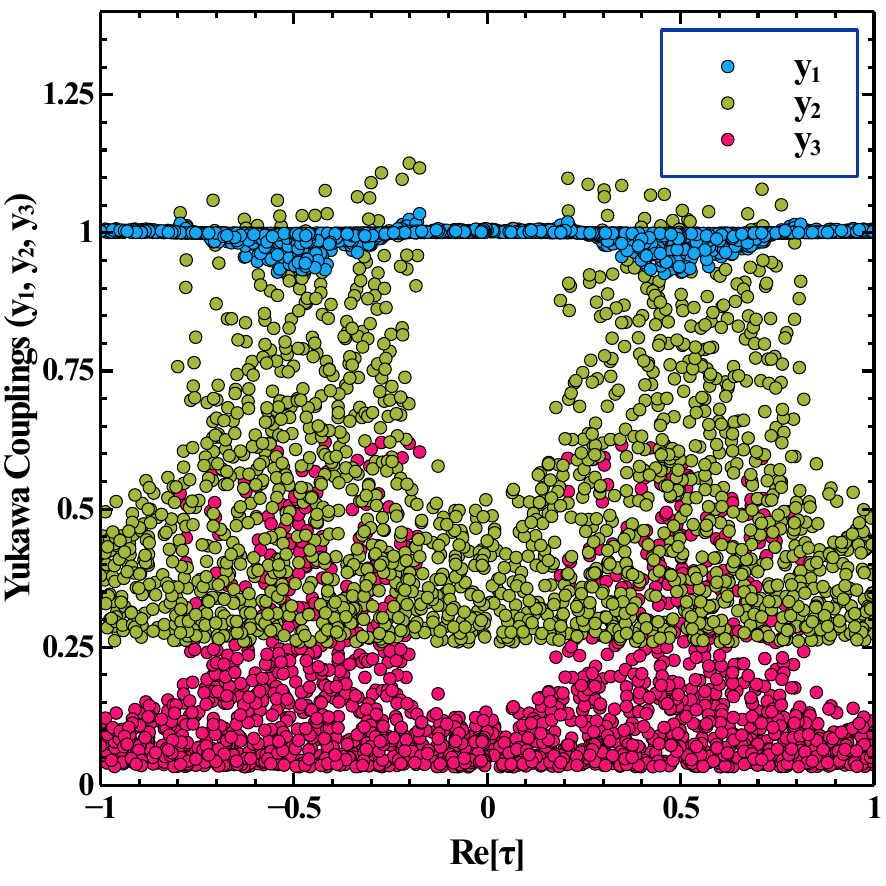}
    \includegraphics[scale=0.45]{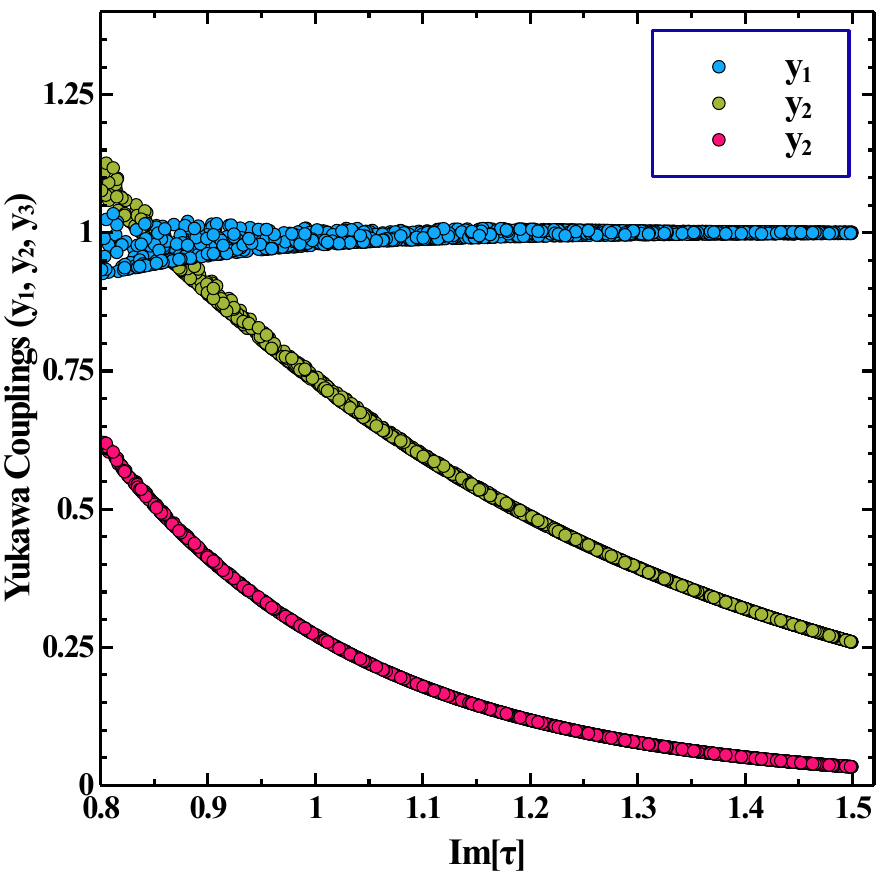}
    \caption{Variation of Yukawa couplings with the real and imaginary components of $\tau$, after going through neutrino scan. }
    \label{fig:fig-3}
\end{figure}
To match the current neutrino oscillation data, we select the following ranges for the model parameters:%\footnote {\color{red} We choose to work on the principal domain of modular symmetry. The principal domain for modular symmetry allows a specific range for the free parameters to satisfy neutrino oscillation data \cite{}. }:
\begin{equation}
\begin{split}
& \text{Re}[\tau] \in [-1.0, 1.0], \quad \text{Im}[\tau] \in [0,2], \quad \{\alpha_{D}, \beta_{D}, \gamma_{D}\} \sim \mathcal{O}(10^{-3}), \\ 
& \alpha_{NS} \sim \mathcal{O}(10^{-1}), \quad \beta_{NS} \sim \mathcal{O}(10^{-4}), \quad m_{\eta_{R}}=m_{\eta_{I}} \in [10^3, 10^4] \, GeV, \\
& k_s \in [1, 10^3]\, GeV, \quad \rho_{0} \in [10^3, 10^6]\, GeV, \quad \beta_{L} \in [0.01,1.0], \quad \zeta_{1,2} \in [0.01,1].
\end{split}
\end{equation}
The input parameters are randomly scanned within the above-specified ranges. The permitted regions are initially filtered based on the \(3\sigma\) limits of solar and atmospheric mass squared differences and mixing angles. From the random scan, the modulus $\tau$ is found to be lie within the range \(0.17 \lesssim |\text{Re}[\tau]| \lesssim 0.80\) for \(0.8 \lesssim \text{Im}[\tau] \lesssim 1.1\) and \(-1.0 \lesssim \text{Re}[\tau] \lesssim 1.0\) for \(1.1 \lesssim \text{Im}[\tau] \lesssim 1.5\) in the case of normal ordering (NO). The correlation between \(\text{Re}[\tau]\) and \(\text{Im}[\tau]\) is shown in the left panel of Fig. \ref{fig:fig-2}. The right panel of Fig. \ref{fig:fig-2} shows the co-relation among the components of the Yukawa couplings $Y$. $y_1$ and $y_3$ hold both the axes, and the colour bar represents the Yukawa coupling $y_2$, respectively.

Fig. \ref{fig:fig-3} shows the variation of Yukawa couplings with Re[\(\tau\)] $(left)$ and Im[\(\tau\)] $(right)$ respectively. The plausible ranges for Yukawa couplings satisfying neutrino data are found to be in the region
\(0.92 \le y_{1} \le 1.03\), \(0.26 \le y_{2} \le 1.13\) and \(0.034 \le y_{3} \le 0.62\).
\begin{figure}[h!]
        \centering
    \includegraphics[scale=0.5]{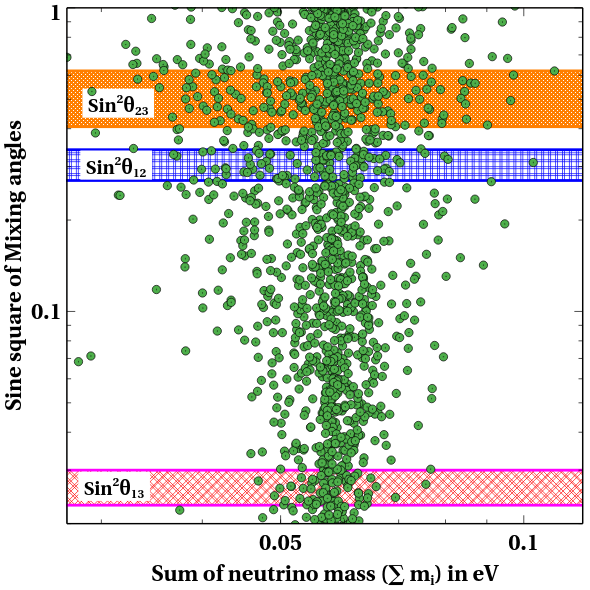}
        \caption{Variation of sine of mixing angles $vs.$ sum of the active neutrino masses in eV. The respective bands represent the sine square of the particular mixing angles.}
        \label{fig:fig-4}
\end{figure}

Fig. \ref{fig:fig-4} illustrates the variation in the sum of total neutrino masses concerning the mixing angles, constrained within the 3\(\sigma\) regions. The sum of the neutrino masses is found to be below the upper bound $\sum m_{\nu} < 0.12$\, eV provided by Planck 2018\cite{Planck:2018vyg} satisfying all the three mixing angles within their current 3$\sigma$ range.

%%%%%%%%%%%%%%%%%%%%%%%%%%%%%%%%%%%%%%%%%%%%%%%%%%%%%%%%%%
\begin{figure}[h!]
    \centering
        \centering
        \includegraphics[scale=0.45]{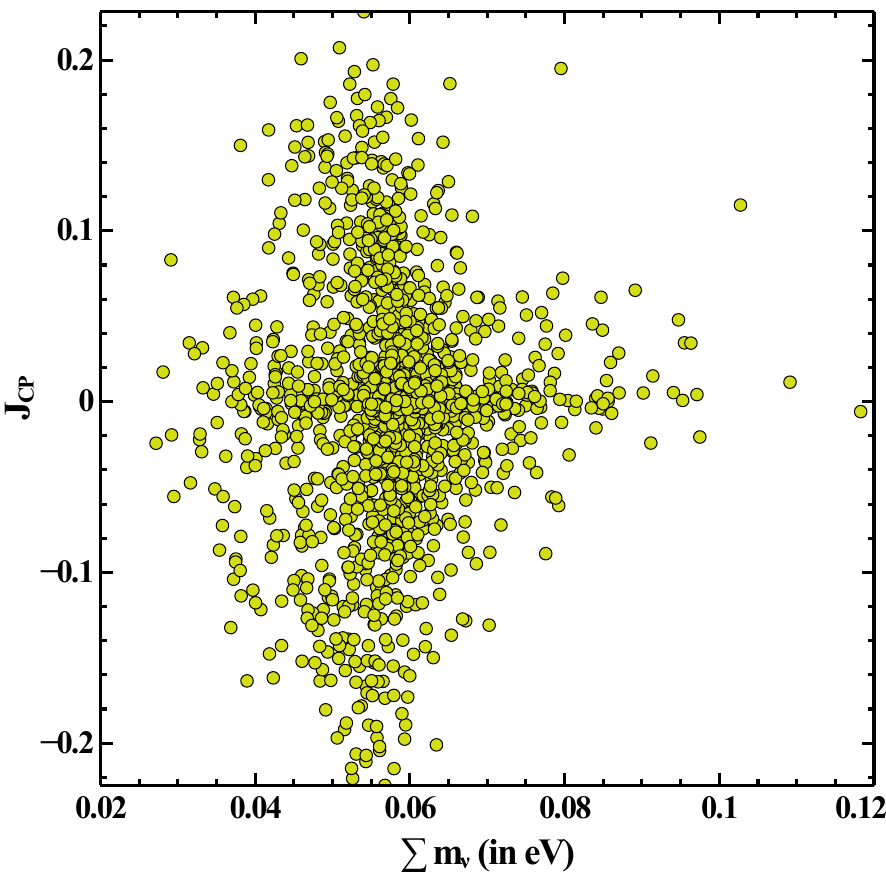}
        \includegraphics[scale=0.45]{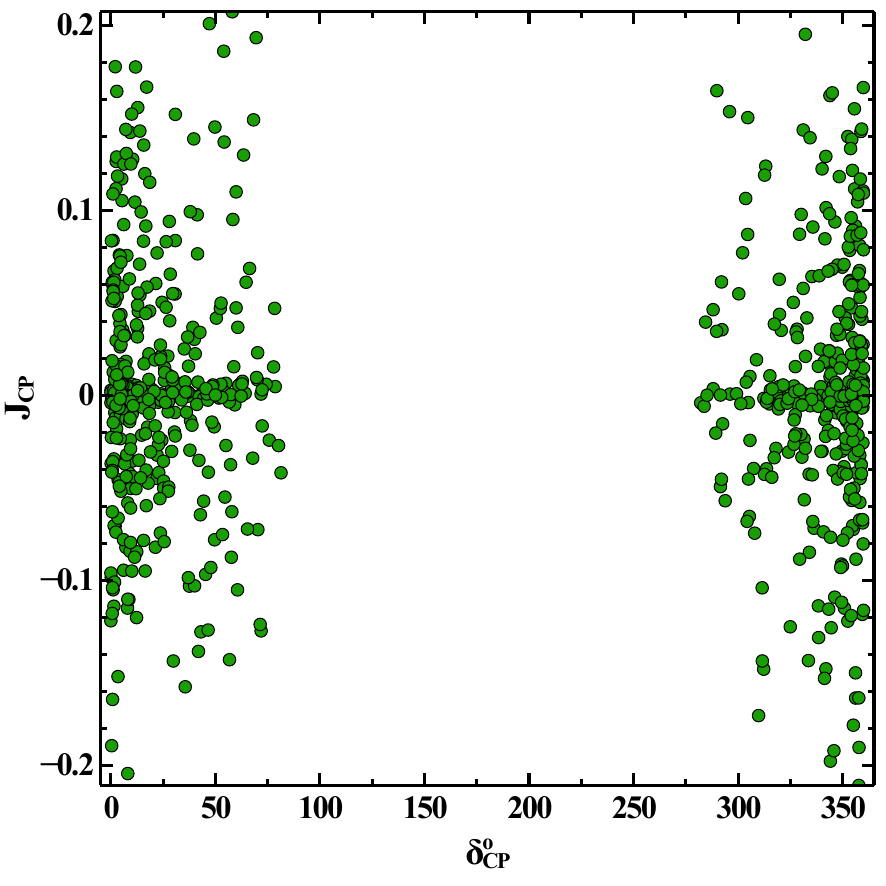}
        \caption{($Left$) Correlation between $J_{CP}$ and sum of neutrino masses $(\sum m_{\nu})$. ($Right$) Variation of $J_{CP}$ with $\delta_{CP}^{o}$.}
        \label{fig:fig-5}
\end{figure}
\begin{figure}[h!]
        \centering
        \includegraphics[scale=0.45]{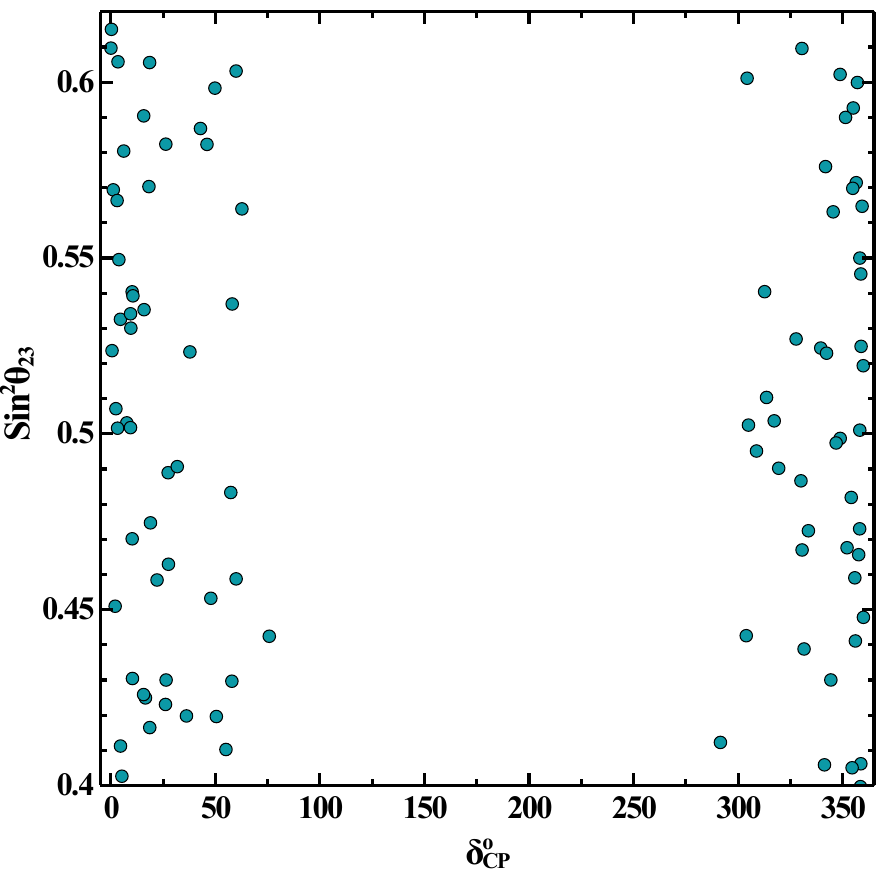}
        \includegraphics[scale=0.45]{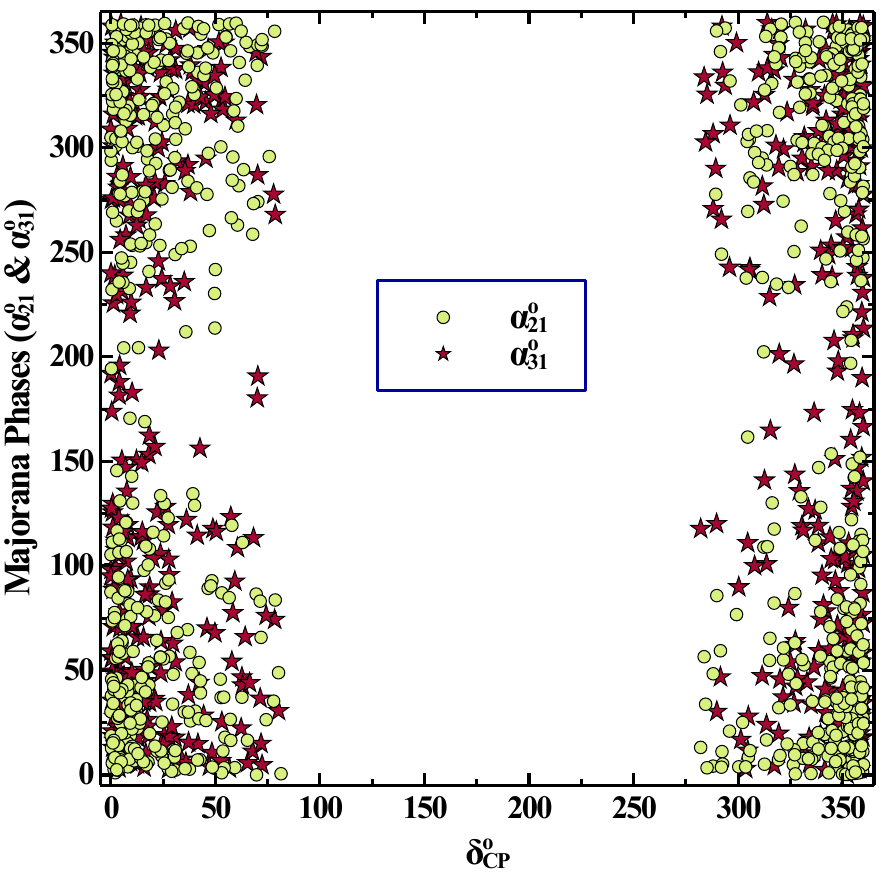}
        \caption{($Left$)Correlation between the Direc CP phase with the Sine square of the atmospheric mixing angle. ($Right$)Interdependence of Majorna phases (\(\alpha_{21}\) \& \(\alpha_{31}\)) with the Dirac CP phase.}
        \label{fig:fig-6}
\end{figure}

%%%%%%%%%%%%%%%%%%%%%%%%%%%%%%%%%%%%%%%%%%%%%%%%%%%%%%%%%%%%%%%
The left panel of Fig. \ref{fig:fig-5} provides insight into the interdependence of Jarlskog CP invariant with the sum of active neutrino masses and its value found to be in the range \(-0.22\lesssim J_{CP}\lesssim 0.22\). Also the variation of $J_{CP}$ with $\delta_{CP}^{o}$ is shown in the right panel of Fig. \ref{fig:fig-5}. The correlation between \(\delta_{CP}^{o}\) and \(\sin^2{\theta_{23}}\) is shown in the left panel of Fig. \ref{fig:fig-6}. The Dirac CP phase is found to be in the region \(\delta_{CP}^{o}\in[0^{\circ},77^{\circ}]\) and \(\delta_{CP}^{o}\in[278^{\circ},360^{\circ}]\). In the right panel of Fig. \ref{fig:fig-6}, we show correlation of Majorana phases \(\alpha_{21}\) and \(\alpha_{31}\) with Dirac phase \(\delta_{CP}^{o}\). 

\begin{figure}[h!]
    \centering
     \includegraphics[width=0.5\textwidth]{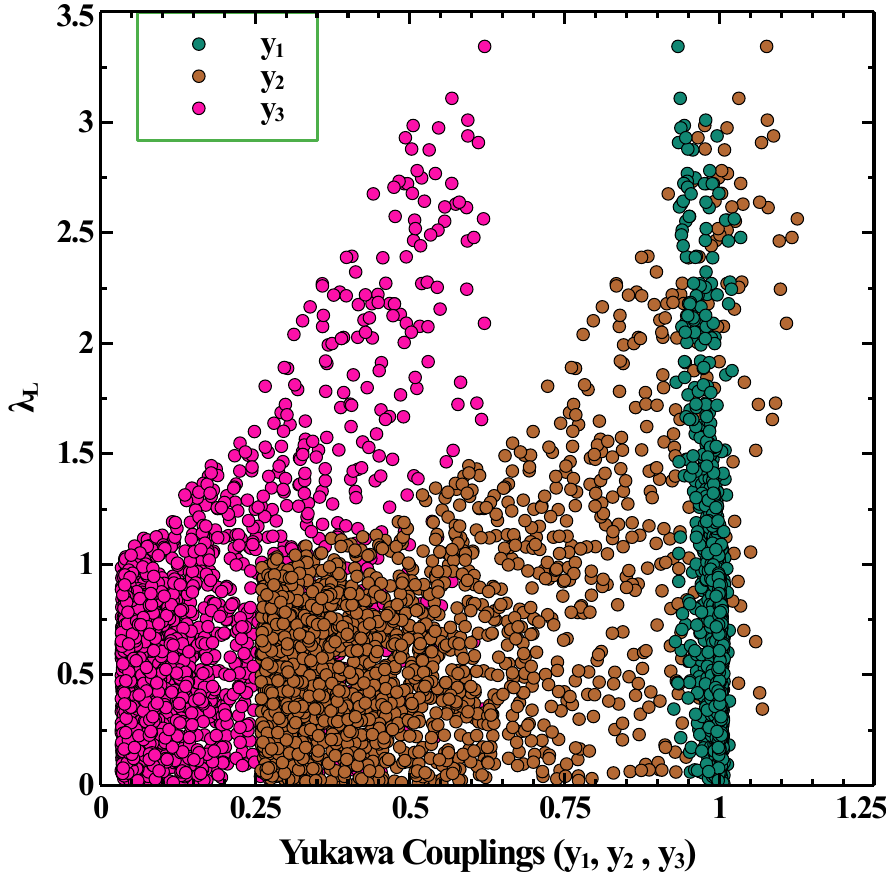}
        \caption{Correlation among the Yukawa couplings $y_{1}, y_{2}, y_{3}$ with the Higgs portal quartic coupling $\lambda_{L}=(\zeta_2\lambda_\eta')$ .}
        \label{fig:fig-7}
\end{figure}

Hence, from the neutrino scan, we find the parameter space for the modulus $\tau$. The same $\tau$ is deciding the fate of the Yukawa as well as the quartic couplings. We define the term $\zeta_{2}\lambda_{\eta}^{'}$ in Eq. \eqref{pot1} as $\lambda_{L}$, which we will be using later in dark matter direct detection section. We show the correlation between $\lambda_{L}$ and the Yukawa couplings $y_{1}, y_{2}, y_{3}$ in Fig. \ref{fig:fig-7}. In the dark matter analysis, we will explicitly use the parameterized $\lambda_L$. It is to be noted that the $\lambda_{\eta}^{'}$ is a $A_{4}$ singlet with $k_{I}=6$, which is expressed in terms of $A_{4}$ triplet with $k_{I}=2$.

%%%%%%%%%%%%%%%%%%%%%%%%%%%%%%%%%%%%%%%%%%%%%%%%%%%%%%%%%%%%%%%
%%%%%%%%%%%%  Fermionic Dark Matter  %%%%%%%%%%%%%%%%%%%
\subsection{Fermionic Dark Matter}
In this section, we will study the possibility of a single component viable dark matter candidate. The odd modular weight is analogous to $Z_2$ discrete charge to stabilize the DM candidate. In our model, among the stable candidates, the fermion \(f\) being the lightest, can serve the purpose of a dark matter. We implemented the model using the {\tt Feynrules} package \cite{Alloul:2013bka} and then retrieved the results with the {\tt micrOMEGAs 6.0} \cite{Alguero:2023zol}. A numerous studies have been carried out with fermionic dark matter in extended scotogenic models \cite{Chun:2023vbh, Karan:2023adm, Borah:2024gql, Borah:2022enh,Ganguly:2023jml,Borah:2023hqw}. Earlier results show the dominance of the $t-$channel annihilation processes over other channels. In our case also, $\eta$ scalar mediated $ff\rightarrow S_{Li}S_{Li}$, and vice-versa processes will be the possible $t-$channel annihilation processes. 
It is seen from equation \eqref{lmu} that the Yukawa couplings involved in the dark matter analysis are also involved in neutrino mass generation. To proceed with the numerical analysis, we used the constrained parameter space allowed from the neutrino scan to find the parameter space satisfied by the current relic density of the DM candidate within the current $3\sigma$ range ($\Omega h^2=0.12\pm0.001$ \cite{Planck:2018vyg}.)

The value of the Yukawa couplings constrained from the neutrino scan is found to be in the following range: \(h_{1}\in[0.0088,1.06], h_{2}\in[0.003,0.6], h_{3}\in[0.0004,0.24]\). Again the model has six doubly degenerate Majorna neutrinos, which masses are found to be in the range \(M_{N_{R1}}=M_{S_{L1}}\in[1.67, 203] \, TeV, M_{N_{R2}}=M_{S_{L2}}\in[2.49,353] \, TeV, M_{N_{R3}}=M_{S_{L3}}\in[4.04,417] \, TeV\) from Eq. \eqref{eq:LFV data}. The singlet scalar fermion \(f\) mass is found to be in the range \(M_{f}\in[1.5, 3500] \, GeV\). Therefore, being the lightest among all, $f$ serves as the potential dark matter candidate in our study and we will be using the notation $M_{DM}$ from beyond. 

\begin{figure}[h!]
        \centering        \includegraphics[width=0.5\linewidth]{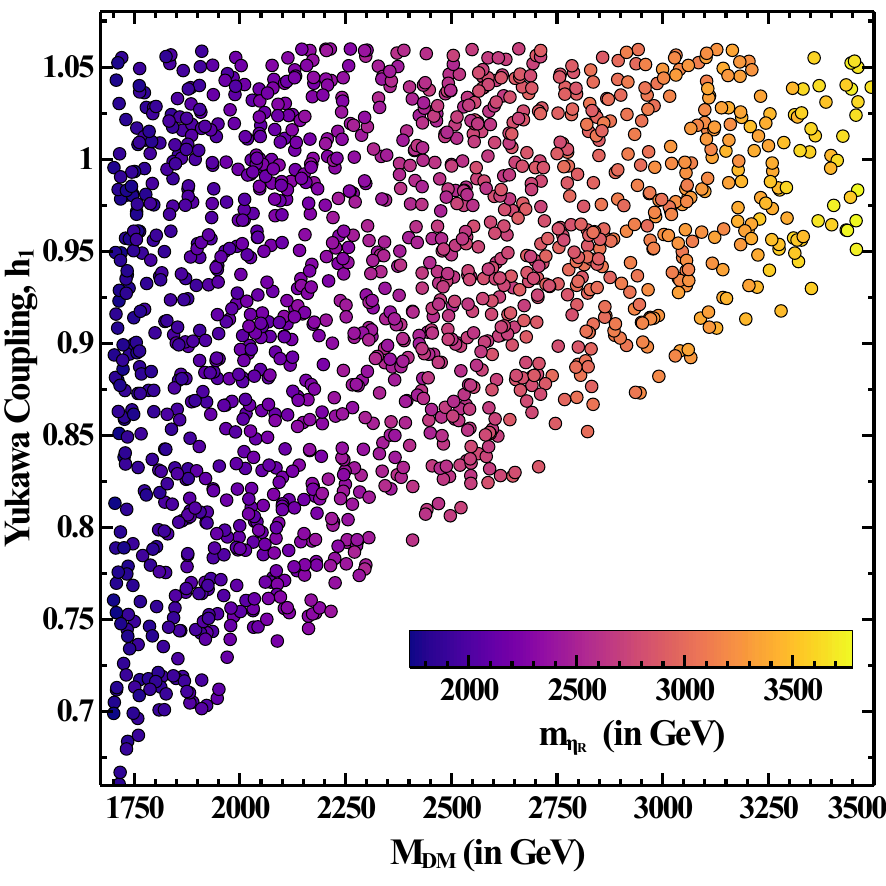}
        \caption{Allowed parameter space for dark matter relic density satisfied points at 3$\sigma$ C.L., in Yukawa coupling vs dark matter mass. The colour bar shows the real singlet scalar mass within the allowed range. }
        \label{dmrelic}
\end{figure}

The parameter space allowed by the DM scan is shown in Fig. \ref{dmrelic}. Since there is a mass hierarchy in the fermion masses (among $S_{L1},S_{L2},S_{L3}$), and constrain in the Yukawa coupling, the only relevant channel giving correct DM relic is via the $\eta_R$ mediated $ff\leftrightarrow S_{L1}S_{L1}$ channel. The masses for the scalar mediator and the fermions start from $M_{S_{L1}}=1.67$ TeV, allowing relic density for the DM candidate in the correct range with slight mass splitting among them. Therefore, the DM relic allowed parameter space starts from $M_{DM}=1.68$ TeV. The larger mass of the other singlet fermions leads to an overabundance of relic density for $S_{L2}$ or $S_{L3}$, restricting the parameter space up to $M_{DM}=3.55$ TeV. One can also see a constrained Yukawa coupling while satisfying 3$\sigma$ relic density for the whole dark matter mass from Fig. \ref{dmrelic}. This pattern of larger Yukawa demand for larger $M_{DM}$ is also justified as the $t$-channel processes are directly proportional to the square of the Yukawas. The mediator mass variation is shown in the colour bar.

%\begin{table}[h!]
 %   \centering
 %   \begin{tabular}{||c|c||}
  %  \hline
   % \hline
    %       Parameter & Chosen range \\
   %      \hline
    %       Singlet fermion $(f)$  & $M_{DM}=[1.5,3500] \, GeV$ \\
     %     Yukawa Couplings & $h_{1} = [0.0088,1.06], h_{2}=[0.003,0.6]$,  $h_{3}=[0.0004,0.24]$\\
      %     Majorana fermions & $M_{S_{L1}}=[1.67,203] \,TeV, M_{S_{L2}}=[2.49,353] \, TeV,\, M_{S_{L3}}=[4.04,417]\,TeV$  \\
       %   Scalar masses $(\eta_{R},\eta_{I})$ & $M_{\eta_{R}}= M_{\eta_{I}}=[1.7,10]\,TeV$  \\
    %\hline
    %\end{tabular}
    %\caption{Parameter space allowed by neutrino oscillation, used in dark matter scan.}
    %\label{tab:dm}
%\end{table}

%%%%%%%%%%%%%%%%%%%   Detections    %%%%%%%%%%%%%%%%%%%%%%%%%%

\section{Detection possibilities}\label{sec4}

%%%%%%%%%%%%%% Neutrinoless double beta decay %%%%%%%%%%%%%%%%%%%%%

\subsection{Neutrinoless Double Beta Decay ($0 \nu \beta \beta$)}
\begin{figure}[h!]
    \centering
        \centering
        \includegraphics[scale=0.5]{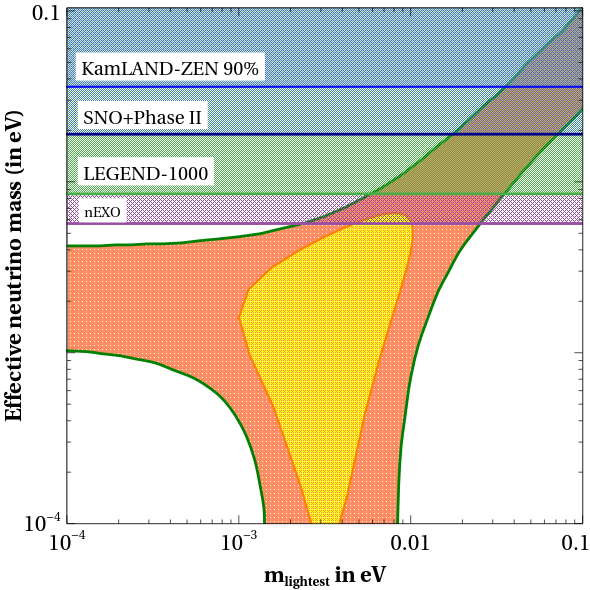}
        \caption{Effective neutrino mass $|m_{ee}|$ as a function of lightest neutrino mass ($m_{1}$). The orange-shaded region represents the currently allowed region for effective neutrino mass in NO. The model prediction from our study finds the yellow region satisfying all neutrino parameters.}
        \label{fig:fig-NDBD}
\end{figure}

Effective neutrino mass $|m_{ee}|$ is one of the important parameters in establishing the Majorana nature of the neutrinos. The half-life of $0 \nu \beta \beta$ is proportional to $|m_{ee}|^2$, where $|m_{ee}|$ contains the information about the absolute mass eigenvalues and the mixing angles along with the phases of the PMNS matrix as given by:
\begin{equation}
    |m_{ee}|=|m_{1}\cos^{2}{\theta_{12}}\cos^{2}{\theta_{13}} + m_{2}\sin^{2}{\theta_{12}}\cos^{2}{\theta_{13}}e^{i \alpha_{21}} + m_{3}\sin^{2}{\theta_{13}}e^{i(\alpha_{31}-2\delta_{CP})}|
   \label{NDBD} 
\end{equation}

The current experimental limit from KamLAND-Zen \cite{KamLAND-Zen:2022tow} experiment is $(36-156)$ meV. The sensitivity limits on $|m_{ee}|$ projected for future experiment SNO + Phase II \cite{SNO:2015wyx} is $(19-46)$ meV, LEGEND-1000 \cite{LEGEND:2021bnm} predicted limits within $(8.5-19.4)$ meV while nEXO \cite{nEXO:2017nam} limits lie within $(5.7-17.7)$ meV at $90\%$ C.L. The sensitivity limits are shown as mentioned in the figure. In Fig. \ref{fig:fig-NDBD}, the orange-shaded region shows the currently allowed neutrino parameter space for the effective mass within 3$\sigma$ C.L. and the yellow contour represents the statistically predicted region by our model. The theoretical predictions from our model constrain the lightest neutrino mass to lie within the range of $\mathcal{O}(10^{-2})$ eV. In this context, both Majorana and Dirac CP phases have a relatively minor impact on limiting the lightest neutrino mass. However, a strong correlation is observed between the lightest neutrino mass (and consequently the sum of neutrino masses) and the neutrino mixing angles, as clearly illustrated in Figure \ref{fig:fig-4}. Therefore, our model keeps a stronghold from the experimental perspective to be probed in near future experiments. % In our model, the variation of $|m_{ee}|$ with the lightest neutrino mass is shown in Fig. \ref{fig:fig-NDBD} in the yellow colour region. Our model predicted region is shown in the yellow contour region, which is within the current 3$\sigma$ range satisfied by neutrino oscillation parameters. 
%%%%%%%%%%%%%%%%%%%%%%%%%%%%%%%%%%%%%%%%%%%%%%%%%%%%%%%%%%%%%%%%
%%%%%% Lepton Flavour Violation %%%%%%%%%%%%%%%%%%%%%%%%%%%%

\subsection{Lepton Flavour Violation}
The search for the lepton flavour violation is exceptionally crucial in the pursuit of new physics beyond the Standard Model (SM). Lepton flavor-violating decays such as \(\mu \rightarrow e + \gamma\), \(\mu \rightarrow eee\), and \(\mu \rightarrow e\) conversion in nuclei, which are suppressed in the Standard Model by the GIM mechanism, could have significant contributions in the current model \cite{Calibbi:2017uvl, Ardu:2022sbt}. 

\begin{figure}[h!]
    \centering
     \includegraphics[width=0.5\textwidth]{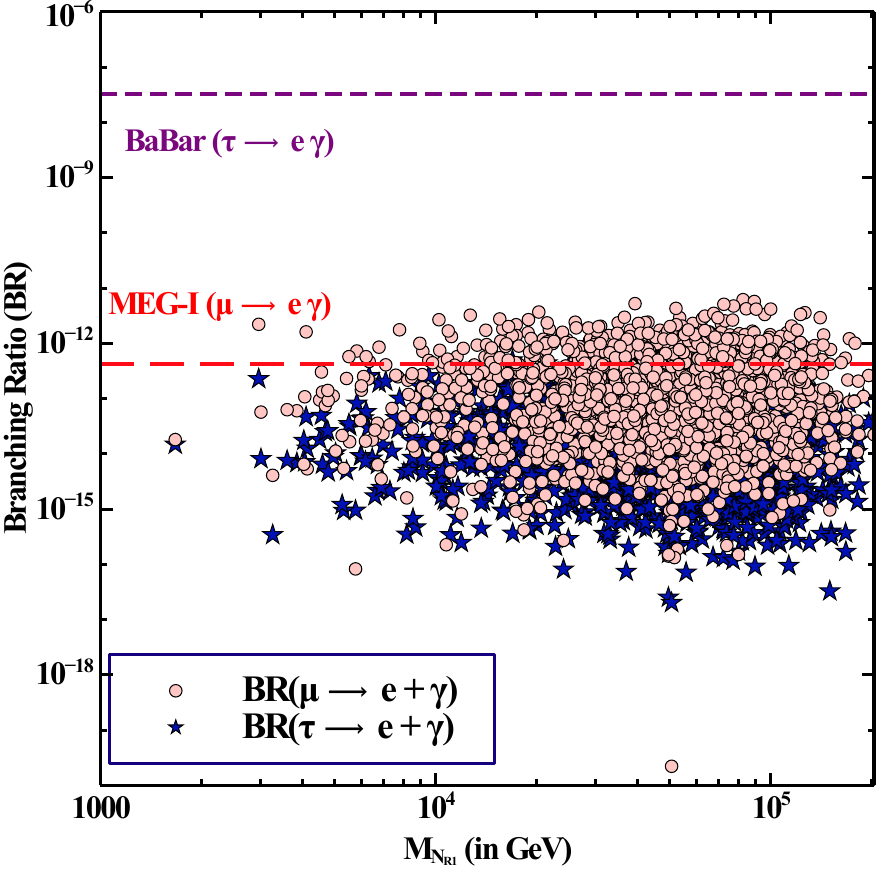}
        \caption{Relationship between \(M_{N_{R1}}\) and branching ratio (BR) of different LFV processes.}
        \label{fig:fig-13}
\end{figure}
The matrix \(\mathbb{M}\) in Eq. \eqref{eq:M matrix} can be diagonalised by the \(9\times9\) unitary matrix U, which is given by:
\begin{equation}
    U^{T}\mathbb{M}U = diag(m_{1}, m_{2}, m_{3}, M_{N_{R1}}, M_{S_{L1}}, M_{N_{R2}}, M_{S_{L2}}, M_{N_{R3}}, M_{S_{L3}})
\label{eq:LFV data}    
\end{equation}
where $m_1, m_2, m_3$ are the three light neutrino masses and the masses of heavy neutrinos are $M_{N_{R1}}, M_{S_{L1}}, M_{N_{R2}}, M_{S_{L2}}, M_{N_{R3}}, M_{S_{L3}}$. In our model, lepton flavour violating decays $l_{i}\rightarrow l_{j}\gamma$ can happen by exchanging heavy fermions at the one-loop level. This is due to the mixing between light and heavy fermions. The dominant one-loop contribution to the branching ratios for these decays is described by the dominant term given as\cite{Chakraborty:2021azg, Deppisch:2004fa, Forero:2011pc, Chekkal:2017eka, Ilakovac:1994kj}:
\begin{equation}
    BR(l_{i}\longrightarrow l_{j}\gamma)=\frac{\alpha^{3}\sin^2{\theta_{W}}}{256\pi^{2}}\left(\frac{M_{l_{i}}}{M_{W}}\right)^{4}\frac{M_{l_{i}}}{\Gamma_{l_{i}}}|G_{ij}|^{2}
\end{equation}
where \(G_{ij}\) is loop function whose analytical form is:
\begin{equation}
 \begin{aligned} G_{ij}=\sum_{k}U_{ik}^{*}U_{jk}G_{\gamma}\left(\frac{M_{k}^{2}}{M_{W}^{2}}\right),\text{with}\\
 G_{\gamma}(x)=-\frac{2x^{3}+5x^{2}-x}{4(1-x)^{2}}-\frac{3x^{3}}{2(1-x)^{4}}\ln{x}.
 \end{aligned}      
\end{equation}
Here, \(\alpha\) is the fine structure constant, \(\theta_{W}\) is the Weinberg angle, \(U_{ij}\) is the $ij^{th}$ element of the unitary matrix $U$, \(M_{W}\) is the mass of $W$ boson, \(M_k\)'s $(k=N_{R1},N_{R2},N_{R3})$ are the masses of heavy fermions and \(M_{l_{i}}\) is the mass of decaying lepton. The decay width for muon is given by \cite{Ilakovac:1994kj}:
\begin{equation}
    \Gamma_{\mu}=\frac{G_{F}^{2}M_{\mu}^{2}}{192\pi^{3}}\left(1-8\frac{M_{e}^{2}}{M_{\mu}^{2}}\right)\left[1+\frac{\alpha}{2\pi}\left(\frac{25}{4}-\pi^{2}\right)\right]
\end{equation}
where \(G_{F}\), \(M_{e}\) and \(M_{\mu}\) are Fermi constant, electron mass and muon mass, respectively. The decay width of \(\tau\) is \(2.267\times 10^{-12}\) GeV \cite{Chakraborty:2021azg}. Fig. \ref{fig:fig-13} shows the branching ratio for the two processes as denoted by the key text. The red horizontal line stands for MEG-I collaboration bound \(4.2 \times 10^{-13}\) \cite{MEG:2016leq,Baldini:2013ke} for the BR$(\mu\rightarrow e\gamma)$. The purple horizontal line represents the experimental upper bound \(3.3 \times 10^{-8}\) set by BaBar collaboration\cite{BaBar:2009hkt} for the $\tau$-decay processes. 
%%%%%%%%%%%%%%%%%%%%%%%%%%%%%%%%%%%%%%%%%%%%%%%%%%%%%%%%%%%%%%%%%%%%%%%%%
%%%%%%%%%%%%%%%% DM Direct Detection %%%%%%%%%%%%%%%%%%%%%%%%%%%%%%%

\subsection{Direct detection}

\begin{figure}
    \centering
    \includegraphics[scale=0.8]{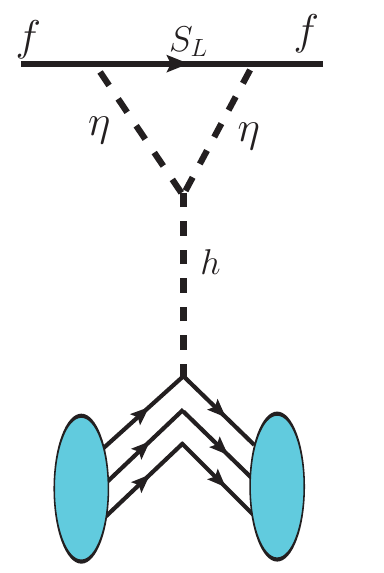}     \includegraphics[scale=0.5]{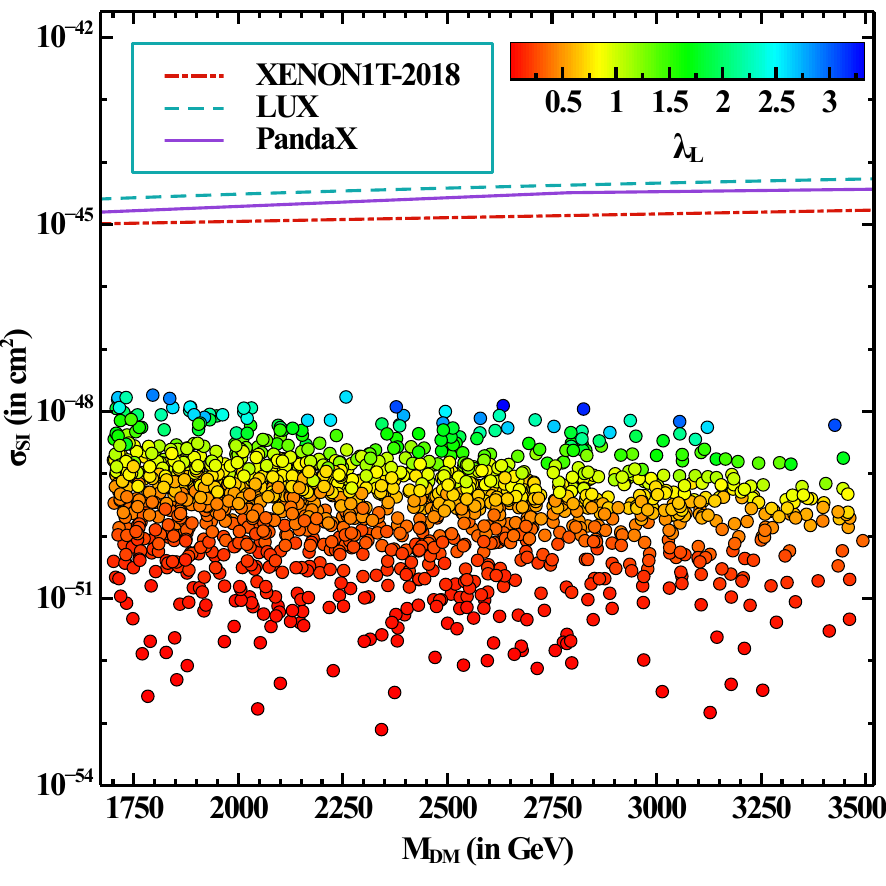}
    \caption{({\it Left})Dark matter contribution to spin-independent direct detection via one-loop process. ({\it Right}) Spin-independent cross-section of dark matter $vs.$ dark matter mass in GeV. The colour bar shows the quartic coupling associated with the scalar vertex. All the points satisfy the relic density of DM at 3$\sigma$ C.L.}
    \label{dd1}
\end{figure}
We extend our search for DM via the direct detection prospects of our modular setup. Even though our singlet fermion DM does not have any tree-level scattering process off nucleons, we can still have a direct detection cross-section as shown in the left panel of Fig. \ref{dd1} at one-loop level via the scalar doublet. The contribution to the spin-independent scattering cross-section for the dark matter-nucleon scattering is given by
\begin{equation}
    \sigma_{\rm SI} = \left(\frac{m_n}{v}\right)^2 \frac{(\mu_{\rm DM-n})^2~ g_{\overline{f} f h}^2}{\pi M_{h}^4} f_{n}^2,
\end{equation}
where, $\mu_{\rm DM-n}=\frac{M_{\rm DM} m_n}{M_{\rm DM}+m_n}$ is the reduced mass of DM-nucleon system with $m_n$ being the mass of the nucleon and  $f_n=0.3$ \cite{Cline:2013gha}. The parameter $ g_{\overline{{f}} f h}$ is the  effective coupling between DM and Higgs, which can be expressed as,
\begin{equation}
    g_{\overline{{f}} f h}= \frac{i}{16 \pi^2} \times \frac{h_{1} \lambda_{L} v}{M_{\rm DM}}\times \left[1+\left(\frac{M_{\eta}^2}{M_{\rm DM}^2} -1\right)\ln\left(1-\frac{M_{\rm DM}^2}{M_{\eta}^2} \right) \right].
\end{equation}
In the right panel of Fig. \ref{dd1}, we can see that all the DM relic satisfied points are well below the recent bounds from the experiments such as LUX\cite{LZ:2022lsv}, XENON1T-2018 \cite{XENON:2019ykp} and PandaX \cite{PandaX:2018wtu} as indicated by the key text with respective horizontal lines. Here, $\lambda_L=\zeta_2\lambda_{\eta}'$ is the Higgs-$\eta$ vertex, whose allowed range\footnote{In modular symmetry, the quartic couplings are not free parameters and they are constrained by their respective modular structures as shown in appendix \ref{sec: Append B}.} is shown in the colour bar.

\section{Conclusion}\label{sec5}

We study an extension of the Standard Model (SM) framework where neutrino masses and dark matter are simultaneously addressed within the scotogenic inverse seesaw framework. Apart from the tree level contribution to the neutrino mass generation, the radiative generation of the Majorana mass term establishes a connection between the mechanism of neutrino mass generation and the dark matter (DM) candidate. To implement this scenario, we adopt a modular symmetry framework based on the $A_4$ flavour symmetry. The odd modular charge of certain fields ensures the stability of the DM candidate, while the specific modular weights of the couplings result in a distinctive flavour structure. 

The model exhibits rich neutrino phenomenology, including strong correlations among parameters and observables. The results yield constrained regions in the neutrino mixing angles, which restrict the sum of the active neutrino masses to the order of  $\mathcal{O}(10^{-2})$ eV, which is consistent with the Jarlskog invariant for CP violation. This result also leads to a constrained region for the lightest neutrino mass, as inferred from the effective mass calculation in the neutrinoless double beta decay ($0\nu\beta\beta$) analysis. 

While the fermionic DM scenario itself is not entirely novel, the key distinction of our model from previous studies lies in its predictive modular framework, where all couplings relevant to DM are functions of a single complex modulus parameter $\tau$. This not only links DM properties to the neutrino sector but also significantly reduces the model's arbitrariness.

%From the dark matter perspective, while we do not claim that the fermionic DM scenario itself is entirely novel, our framework uniquely links the viable DM parameter space with the origin of neutrino mass, offering a theoretically motivated connection between neutrino physics and dark matter phenomenology. Notably, all the couplings involved in the DM analysis have a modular form and are functions of a single complex parameter $\tau$, adding further theoretical consistency and predictivity to the model.

%From an experimental perspective, we investigate the contributions of our model to processes such as neutrinoless double beta decay ($0\nu\beta\beta$) and charged lepton flavour violation (e.g., $\mu\rightarrow e\gamma$ and $\tau\rightarrow e\gamma$). The fermionic DM candidate achieves the observed thermal relic abundance at the TeV scale, primarily governed by the Yukawa couplings. To ensure the potential detectability of the DM candidate, we also examine its spin-independent direct detection cross-section in the context of future experiments.
On the phenomenological side, we investigate signatures of the model in neutrinoless double beta decay ($0\nu\beta\beta$) and charged lepton flavour violation (e.g., $\mu \rightarrow e\gamma$, $\tau \rightarrow e\gamma$). The fermionic DM candidate achieves the correct relic abundance at the TeV scale, driven by Yukawa interactions. Additionally, we examine the spin-independent DM-nucleon scattering cross-section to assess the model’s compatibility with upcoming direct detection experiments.

%%%%%%%%%%%%%%%%%%    Acknowledgement %%%%%%%%%%%%%%%%%%%%%%%%%%%%

\section*{Acknowledgement}
GP would like to acknowledge CSIR-HRDG for the financial support received in the form of JRF fellowship (09/0796(16046)/2022-EMR-I). 

%%%%%%%%%%%%%%%%%%%%%     Appendix     %%%%%%%%%%%%%%%%%%%%%%%%
%%%%%%%%%%%      Appendix A         %%%%%%%%%%%%%%%%%%%%%%%%%%%%
\appendix

\renewcommand{\theequation}{A\arabic{equation}} % Prefix equations with "A"

\setcounter{equation}{0} % Reset equation counter
%\begin{center}
\section{$A_4$ Group}
$A_4$ is the symmetry group of the tetrahedron and the group of even permutations of four objects. It therefore has $4!/2 = 12$ elements. It can be seen that all twelve elements can be obtained by repeatedly multiplying the two generators, $S = (14)(23)$ and $T = (123)$. These satisfy the relations:\\
\begin{equation*}
    S^2=(ST)^3=T^3=1.    
\end{equation*}
 $A_4$ has four inequivalent representations: three of dimension one, 1, $1'$ and $1''$ and one of dimension 3. The product rule for $A_4$ group  is given as follow:
 \begin{equation}  
\begin{split}
   & 1\otimes 1=1\\
   & 1'\otimes 1''=1\\
   & 1'\otimes 1'=1''\\
   & 1^{(')('')}\otimes 3=3\\
   & 3\otimes 3= 1\oplus 1'\oplus 1''\oplus 3_{S}\oplus 3_{A}
\end{split}    
\end{equation}
where, $3_{S(A)}$ represents symmetric (anti-symmetric) combination. There are two basis for triplet multiplication: Ma-Rajasekaran basis and Altarelli-Feruglio basis.
 In the Altarelli-Feruglio (AF) basis the generator $T$ is diagonal.
 \begin{equation}
 \begin{split}
     T = \begin{pmatrix}
         1 & 0 & 0 \\
         0 & \omega & 0 \\
         0 & 0 & \omega^2 \\
     \end{pmatrix}, \,
     S = \frac{1}{3} \begin{pmatrix}
         -1 & 2 & 2 \\
         2 & -1 & 2 \\
         2 & 2 & -1
     \end{pmatrix} .
 \end{split}
 \end{equation}
 Here, $\omega$ is the cubic root of unity.\\
 The multiplication rule for $3\otimes 3$ is given for triplets $a=(a_1, a_2, a_3)$ and $b=(b_1, b_2, b_3)$ in the AF basis as follows:
  \begin{equation}
 \begin{split}
 & 1_{AF}: (ab)_1 = a_1 b_1 + a_3 b_2 + a_2 b_3,\\
 & 1'_{AF}: (ab)_{1'} = a_1 b_2 + a_2 b_1 + a_3 b_3,\\
 & 1''_{AF}: (ab)_{1''} = a_1 b_3 + a_3 b_1 + a_2 b_2,\\
 & 3_{AF}^s: \frac{1}{\sqrt{3}}(2a_1 b_1 - a_2 b_3 - a_3 b_2, 2a_3 b_3 - a_1 b_2 - a_2 b_1, 2a_2 b_2 - a_3 b_1 - a_1 b_3)\\
 & 3_{AF}^A: (a_2 b_3 - a_3 b_2, a_1 b_2 - a_2 b_1, a_3 b_1 - a_1 b_3)
 \end{split}    
 \end{equation}
%%%%%%%%%%%%%%%%%     Appendix B %%%%%%%%%%%%%%%%%%%%%%%%%%
\renewcommand{\theequation}{B\arabic{equation}} % Prefix equations with "B"
\setcounter{equation}{0} % Reset equation counter
\section{ Modular forms of Yukawa couplings}
\label{sec: Append B}

$\bar{\Gamma}$ is the modular group that attains a linear fractional transformation $\gamma$ which acts on modulus $\tau$ linked to the upper-half complex plane whose transformation is given by:
\begin{equation}
\gamma \longrightarrow \gamma \tau = \frac{a\tau + b}{c\tau + d},
\end{equation}
where $a, b, c, d \in  \mathbb{Z}$ and $ad-bc = 1$, $Im[\tau]>0$, where it is isomorphic to the transformation $PSL(2, \mathbb{Z}) = SL(2, \mathbb{Z})/\{I, -I\}$. The S and T transformation helps in generating the modular transformation defined by:
\begin{equation}
    S: \tau \longrightarrow -\frac{1}{\tau}, \hspace{1.5cm} T: \tau \longrightarrow \tau + 1,
\end{equation}
and hence the algebraic relations so satisfied are as follows,
\begin{equation}
    S^2 = \mathbb{I}, \hspace{2.5cm} (ST)^3 = \mathbb{I}.
\end{equation}
Here, series of groups are introduced, $\Gamma(N) (N = 1, 2, 3, .....)$ and defined as
\begin{equation}
    \Gamma(N) =\Bigg \{\begin{pmatrix}
        a & b \\
        c & d
    \end{pmatrix} \in SL(2,\mathbb{Z}), \begin{pmatrix}
        a & b \\
        c & d 
    \end{pmatrix} = \begin{pmatrix}
        1 & 0 \\
        0 & 1
    \end{pmatrix}(\text{mod} N)\Bigg\}
\end{equation}
Definition of $\bar{\Gamma}(2) \equiv \Gamma(2)/\{I, -I\}$ for N=2. Since $-I$ is not associated with $\Gamma(N)$ for $N>2$ case, one can have $\bar{\Gamma}(N)=\Gamma(N)$, which are infinite normal subgroups of $\bar{\Gamma}$ known as principal congruence subgroups. Quotient groups come from the finite modular group, defined as $\Gamma_N = \bar{\Gamma}/\bar{\Gamma}(N)$. The
imposition of $T^N = \mathbb{I}$ is done for these finite groups $\Gamma_N$. Thus, the groups $\Gamma_N (N = 2, 3, 4, 5)$ are isomorphic to $S_3$, $A_4$, $S_4$ and $A_5$, respectively. N level modular forms are holomorphic
functions $f(\tau)$ which are transformed under the influence of $\Gamma(N)$ as follows:
\begin{equation}
    f(\gamma\tau)=(c\tau + d)^{k}f(\tau), \hspace{1.5cm} \gamma \in \Gamma(N)
\end{equation}
where k is the modular weight. 

%%%%%%%%%%%%%%%%%%%%%%%%%%%%%%%%%%%%%%%%%%%%%%%%%%%%%%%%%%%%%%%%

 A field $\phi^{(I)}$ transforms under the modular transformation as:
 \begin{equation}
     \phi^{(I)} \rightarrow (c\tau + d)^{-k_I}\rho^{I}(\gamma)\phi^{(I)}
 \end{equation}
 where $-k_I$ represents the modular weight and $\rho^{(I)}(\gamma)$ signifies an unitary representation matrix of $\gamma \in \Gamma(2)$.

%%%%%%%%%%%%%%%%%%%%%%%%%%%%%%%%%%%%%%%%%%%%%%%%%%%%%%%%%%%%%%%%%%

The scalar field's kinetic term is as follows:
\begin{equation}
    \sum_{I} \frac{|{\partial_{\mu}\phi^{(I)}}|^2}{(-i\tau + i\bar{\tau})^{k_I}}
\end{equation}

%%%%%%%%%%%%%%%%%%%%%%%%%%%%%%%%%%%%%%%%%%%%%%%%%%%%%%%%%%%%%%%%%%%%

The modular forms of the Yukawa coupling $Y = (y_1, y_2, y_3)$ with weight 2, which transforms as a triplet of $A_4$ can be expressed in terms of Dedekind eta-function $\eta(\tau)$ and its derivative:
\begin{equation}
\begin{split}
& y_1(\tau) = \frac{i}{2\pi}\bigg(\frac{\eta'(\tau/3)}{\eta(\tau/3)} + \frac{\eta'((\tau + 1)/3)}{\eta((\tau+1)/3)} + \frac{\eta'((\tau+2)/3)}{\eta((\tau+2)/3)} - \frac{27\eta'(3\tau)}{\eta(3\tau)}\bigg), \\  
& y_2(\tau) = \frac{-i}{\pi}\bigg(\frac{\eta'(\tau/3)}{\eta(\tau/3)} + \omega^2 \frac{\eta'((\tau + 1)/3)}{\eta((\tau+1)/3)} + \omega \frac{\eta'((\tau+2)/3)}{\eta((\tau+2)/3)}\bigg), \\
& y_3(\tau) = \frac{-i}{\pi}\bigg(\frac{\eta'(\tau/3)}{\eta(\tau/3)} + \omega \frac{\eta'((\tau + 1)/3)}{\eta((\tau+1)/3)} + \omega^2 \frac{\eta'((\tau+2)/3)}{\eta((\tau+2)/3)}\bigg), \\
\end{split}    
\end{equation}
The Dedekind eta-function $\eta(\tau)$ is given by:
\begin{equation}
    \eta(\tau) = q^{1/24} \prod_{n=1}^{\infty}(1 - q^n), \hspace{1.5cm} q \equiv e^{i2\pi\tau}
\end{equation}
In the form of $q$-expansion, the modular Yukawa can be expressed as:
\begin{equation}
\begin{split}
 &   y_1(\tau) = 1 + 12q + 36q^2 + 12q^3 + ............., \\
  &  y_2(\tau) = -6q^{1/3}(1 + 7q + 8q^2 + .............), \\
   & y_3(\tau) = -18q^{2/3}(1 + 2q + 5q^2 + .............).
\end{split}    
\end{equation}
From the $q$-expansion, we have the following constraint for modular Yukawa couplings:
\begin{equation}
    y_2^{2} + 2 y_1 y_3 = 0
\end{equation}
Higher modular weight Yukawa couplings can be constructed from weight 2 Yukawa $\mathbf{(Y^{(2)})}$ using the $A_4$ multiplication rule. For modular weight $k = 4$, we have the following five modular forms:
\begin{equation}
\begin{split}
  &  Y_{3}^{(4)} = (y_1^2 - y_2 y_3, y_{3}^{2}-y_1 y_2, y_2^2 - y_1 y_3),\\
   & Y_{1}^{(4)} = y_1^2 + 2 y_2 y_3,\\
   & Y_{1'}^{(4)} = y_3^2 + 2 y_1 y_2
\end{split}    
\end{equation}
For modular weight $k = 6$, there are seven modular forms:
\begin{equation}
\begin{split}
& Y_1^{(6)} = y_1^3 + y_2^3 + y_3^3 - 3y_1y_2y_3,\\
& Y_{3a}^{(6)}=(y_{31},y_{32},y_{33}) = (y_1^3 + 2y_1y_2y_3, y_1^2y_2 + 2 y_2^2y_3, y_1^2y_3 + 2y_3^2y_2),\\
& Y_{3b}^{(6)} = (y_3^3 + 2y_1y_2y_3, y_3^2y_1 + 2 y_1^2y_2, y_3^2y_2 + 2y_2^2y_1)
\end{split}
\label{eq:weight6}
\end{equation} 
For modular weight $k=8$, there are nine modular forms:
\begin{equation}
\begin{split}
& Y_1^{(8)} = (Y_1^{2} + 2Y_2Y_3)^2, \quad Y_{1'}^{(8)} = (Y_1^{2} + 2Y_2Y_3)(Y_3^{2} + 2Y_1Y_2), \quad Y_{1''}^{(8)} = (Y_3^{2} + 2Y_1Y_2)^2,\\
& Y_{3a}^{(8)} = (Y_1^{2} + 2Y_2Y_3)\begin{pmatrix}
    Y_1^{2} - Y_2Y_3\\
    Y_3^{2} - Y_1Y_2\\
    Y_2^{2} - Y_1Y_3
\end{pmatrix}, \quad Y_{3b}^{(8)} = (Y_3^{2} + 2Y_1Y_2)\begin{pmatrix}
    Y_2^{2} - Y_1Y_3\\
    Y_1^{2} - Y_2Y_3\\
    Y_3^{2} - Y_1Y_2
\end{pmatrix}
\end{split}
\end{equation}
For modular weight $k=10$, there are eleven modular forms:
\begin{equation}
\begin{split}
& Y_1^{(10)} = (Y_1^{2} + 2Y_2Y_3)(Y_1^{3} + Y_2^{3} + Y_3^{3} - 3Y_1Y_2Y_3),\\
& Y_{1'}^{(10)} = (Y_3^{2} + 2Y_1Y_2)(Y_1^{3} + Y_2^{3} + Y_3^{3} - 3Y_1Y_2Y_3),\\
& Y_{3a}^{(10)} = (Y_1^{2} + 2Y_2Y_3)^2\begin{pmatrix}
    Y_1\\
    Y_2\\
    Y_3
\end{pmatrix}, \quad Y_{3b}^{(10)} = (Y_3^{2} + 2Y_1Y_2)^2\begin{pmatrix}
    Y_2\\
    Y_3\\
    Y_1
\end{pmatrix},\\
& Y_{3c}^{(10)} = (Y_1^{2} + 2Y_2Y_3)(Y_3^{2} + 2Y_1Y_2)\begin{pmatrix}
    Y_3\\
    Y_1\\
    Y_2
\end{pmatrix}.
\end{split}
\end{equation}
For modular weight $k=12$, there are thirteen modular forms:
\begin{equation}
\begin{split}
& Y_{1a}^{(12)} = (Y_1^{2} + 2Y_2Y_3)^{3}, \quad Y_{1b}^{(12)} = (Y_3^{2} + 2Y_1Y_2)^{3},\\
& Y_{1'}^{(12)} = (Y_1^{2} + 2Y_2Y_3)^{2}(Y_3^{2} + 2Y_1Y_2), \quad Y_{1''}^{(12)} = (Y_1^{2} + 2Y_2Y_3)(Y_3^{2} + 2Y_1Y_2)^{2},\\
& Y_{3a}^{(12)} = 2(Y_1^{2} + 2Y_2Y_3)^{2}\begin{pmatrix}
    Y_1^{2} - Y_2Y_3\\
    Y_3^{2} - Y_1Y_2\\
    Y_2^{2} - Y_1Y_3
\end{pmatrix}, \quad Y_{3b}^{(12)} = -2(Y_3^{2} + 2Y_1Y_2)^{2}\begin{pmatrix}
    Y_3^{2} - Y_1Y_2\\
    Y_2^{2} - Y_1Y_3\\
    Y_1^{2} - Y_2Y_3
\end{pmatrix},\\
& Y_{3c}^{(12)} = (Y_1^{2} + 2Y_2Y_3)(Y_3^{2} + 2Y_1Y_2)\begin{pmatrix}
    Y_2^{2} - Y_1Y_3\\
    Y_1^{2} - Y_2Y_3\\
    Y_3^{2} - Y_1Y_2
\end{pmatrix}.
\end{split}
\label{eq:weight12}
\end{equation}
%%%%%%%%%%%%%%%%%%%%%%%%%%%%%%%%%%%%%%%%%%%%%%%%%%%%%%%%%%%%%%%%%

%\bibliographystyle{JHEP}
\bibstyle{apsrev}
\bibliography{ref.bib}

\begin{thebibliography}{77}
\expandafter\ifx\csname natexlab\endcsname\relax\def\natexlab#1{#1}\fi
\expandafter\ifx\csname bibnamefont\endcsname\relax
  \def\bibnamefont#1{#1}\fi
\expandafter\ifx\csname bibfnamefont\endcsname\relax
  \def\bibfnamefont#1{#1}\fi
\expandafter\ifx\csname citenamefont\endcsname\relax
  \def\citenamefont#1{#1}\fi
\expandafter\ifx\csname url\endcsname\relax
  \def\url#1{\texttt{#1}}\fi
\expandafter\ifx\csname urlprefix\endcsname\relax\def\urlprefix{URL }\fi
\providecommand{\bibinfo}[2]{#2}
\providecommand{\eprint}[2][]{\url{#2}}

\bibitem[{\citenamefont{Giunti and Laveder}(2003)}]{Giunti:2003qt}
\bibinfo{author}{\bibfnamefont{C.}~\bibnamefont{Giunti}} \bibnamefont{and}
  \bibinfo{author}{\bibfnamefont{M.}~\bibnamefont{Laveder}}
  (\bibinfo{year}{2003}), \eprint{hep-ph/0310238}.

\bibitem[{\citenamefont{Tanabashi et~al.}(2018)}]{ParticleDataGroup:2018ovx}
\bibinfo{author}{\bibfnamefont{M.}~\bibnamefont{Tanabashi}}
  \bibnamefont{et~al.} (\bibinfo{collaboration}{Particle Data Group}),
  \bibinfo{journal}{Phys. Rev. D} \textbf{\bibinfo{volume}{98}},
  \bibinfo{pages}{030001} (\bibinfo{year}{2018}).

\bibitem[{\citenamefont{Mohapatra et~al.}(2007)}]{Mohapatra:2005wg}
\bibinfo{author}{\bibfnamefont{R.~N.} \bibnamefont{Mohapatra}}
  \bibnamefont{et~al.}, \bibinfo{journal}{Rept. Prog. Phys.}
  \textbf{\bibinfo{volume}{70}}, \bibinfo{pages}{1757} (\bibinfo{year}{2007}),
  \eprint{hep-ph/0510213}.

\bibitem[{\citenamefont{Aghanim et~al.}(2020)}]{Planck:2018vyg}
\bibinfo{author}{\bibfnamefont{N.}~\bibnamefont{Aghanim}} \bibnamefont{et~al.}
  (\bibinfo{collaboration}{Planck}), \bibinfo{journal}{Astron. Astrophys.}
  \textbf{\bibinfo{volume}{641}}, \bibinfo{pages}{A6} (\bibinfo{year}{2020}),
  \bibinfo{note}{[Erratum: Astron.Astrophys. 652, C4 (2021)]},
  \eprint{1807.06209}.

\bibitem[{\citenamefont{Bennett et~al.}(2013)\citenamefont{Bennett, Larson,
  Weiland, Jarosik, Hinshaw, Odegard, Smith, Hill, Gold, Halpern
  et~al.}}]{bennett2013nine}
\bibinfo{author}{\bibfnamefont{C.~L.} \bibnamefont{Bennett}},
  \bibinfo{author}{\bibfnamefont{D.}~\bibnamefont{Larson}},
  \bibinfo{author}{\bibfnamefont{J.~L.} \bibnamefont{Weiland}},
  \bibinfo{author}{\bibfnamefont{N.}~\bibnamefont{Jarosik}},
  \bibinfo{author}{\bibfnamefont{G.}~\bibnamefont{Hinshaw}},
  \bibinfo{author}{\bibfnamefont{N.}~\bibnamefont{Odegard}},
  \bibinfo{author}{\bibfnamefont{K.}~\bibnamefont{Smith}},
  \bibinfo{author}{\bibfnamefont{R.}~\bibnamefont{Hill}},
  \bibinfo{author}{\bibfnamefont{B.}~\bibnamefont{Gold}},
  \bibinfo{author}{\bibfnamefont{M.}~\bibnamefont{Halpern}},
  \bibnamefont{et~al.}, \bibinfo{journal}{The Astrophysical Journal Supplement
  Series} \textbf{\bibinfo{volume}{208}}, \bibinfo{pages}{20}
  (\bibinfo{year}{2013}).

\bibitem[{\citenamefont{Zwicky}(1937)}]{Zwicky:1937zza}
\bibinfo{author}{\bibfnamefont{F.}~\bibnamefont{Zwicky}},
  \bibinfo{journal}{Astrophys. J.} \textbf{\bibinfo{volume}{86}},
  \bibinfo{pages}{217} (\bibinfo{year}{1937}).

\bibitem[{\citenamefont{Rubin and Ford~Jr}(1970)}]{rubin1970rotation}
\bibinfo{author}{\bibfnamefont{V.~C.} \bibnamefont{Rubin}} \bibnamefont{and}
  \bibinfo{author}{\bibfnamefont{W.~K.} \bibnamefont{Ford~Jr}},
  \bibinfo{journal}{Astrophysical Journal, vol. 159, p. 379}
  \textbf{\bibinfo{volume}{159}}, \bibinfo{pages}{379} (\bibinfo{year}{1970}).

\bibitem[{\citenamefont{Das et~al.}(2017)\citenamefont{Das, Nomura, Okada, and
  Roy}}]{Das:2017ski}
\bibinfo{author}{\bibfnamefont{A.}~\bibnamefont{Das}},
  \bibinfo{author}{\bibfnamefont{T.}~\bibnamefont{Nomura}},
  \bibinfo{author}{\bibfnamefont{H.}~\bibnamefont{Okada}}, \bibnamefont{and}
  \bibinfo{author}{\bibfnamefont{S.}~\bibnamefont{Roy}},
  \bibinfo{journal}{Phys. Rev. D} \textbf{\bibinfo{volume}{96}},
  \bibinfo{pages}{075001} (\bibinfo{year}{2017}), \eprint{1704.02078}.

\bibitem[{\citenamefont{King}(2016)}]{King:2015sfk}
\bibinfo{author}{\bibfnamefont{S.~F.} \bibnamefont{King}},
  \bibinfo{journal}{Nucl. Phys. B} \textbf{\bibinfo{volume}{908}},
  \bibinfo{pages}{456} (\bibinfo{year}{2016}), \eprint{1511.03831}.

\bibitem[{\citenamefont{Barrie et~al.}(2022)\citenamefont{Barrie, Han, and
  Murayama}}]{Barrie:2022cub}
\bibinfo{author}{\bibfnamefont{N.~D.} \bibnamefont{Barrie}},
  \bibinfo{author}{\bibfnamefont{C.}~\bibnamefont{Han}}, \bibnamefont{and}
  \bibinfo{author}{\bibfnamefont{H.}~\bibnamefont{Murayama}},
  \bibinfo{journal}{JHEP} \textbf{\bibinfo{volume}{05}}, \bibinfo{pages}{160}
  (\bibinfo{year}{2022}), \eprint{2204.08202}.

\bibitem[{\citenamefont{Mohapatra}(2004)}]{Mohapatra:2004zh}
\bibinfo{author}{\bibfnamefont{R.~N.} \bibnamefont{Mohapatra}}, in
  \emph{\bibinfo{booktitle}{{SEESAW25: International Conference on the Seesaw
  Mechanism and the Neutrino Mass}}} (\bibinfo{year}{2004}), pp.
  \bibinfo{pages}{29--44}, \eprint{hep-ph/0412379}.

\bibitem[{\citenamefont{Abada et~al.}(2021)\citenamefont{Abada, Bernal,
  Hern\'andez, Marcano, and Piazza}}]{Abada:2021yot}
\bibinfo{author}{\bibfnamefont{A.}~\bibnamefont{Abada}},
  \bibinfo{author}{\bibfnamefont{N.}~\bibnamefont{Bernal}},
  \bibinfo{author}{\bibfnamefont{A.~E.~C.} \bibnamefont{Hern\'andez}},
  \bibinfo{author}{\bibfnamefont{X.}~\bibnamefont{Marcano}}, \bibnamefont{and}
  \bibinfo{author}{\bibfnamefont{G.}~\bibnamefont{Piazza}},
  \bibinfo{journal}{Eur. Phys. J. C} \textbf{\bibinfo{volume}{81}},
  \bibinfo{pages}{758} (\bibinfo{year}{2021}), \eprint{2107.02803}.

\bibitem[{\citenamefont{Dias et~al.}(2012)\citenamefont{Dias, de~S.~Pires,
  Rodrigues~da Silva, and Sampieri}}]{Dias:2012xp}
\bibinfo{author}{\bibfnamefont{A.~G.} \bibnamefont{Dias}},
  \bibinfo{author}{\bibfnamefont{C.~A.} \bibnamefont{de~S.~Pires}},
  \bibinfo{author}{\bibfnamefont{P.~S.} \bibnamefont{Rodrigues~da Silva}},
  \bibnamefont{and} \bibinfo{author}{\bibfnamefont{A.}~\bibnamefont{Sampieri}},
  \bibinfo{journal}{Phys. Rev. D} \textbf{\bibinfo{volume}{86}},
  \bibinfo{pages}{035007} (\bibinfo{year}{2012}), \eprint{1206.2590}.

\bibitem[{\citenamefont{Gautam and Das}(2021)}]{Gautam:2020wsd}
\bibinfo{author}{\bibfnamefont{N.}~\bibnamefont{Gautam}} \bibnamefont{and}
  \bibinfo{author}{\bibfnamefont{M.~K.} \bibnamefont{Das}},
  \bibinfo{journal}{Int. J. Mod. Phys. A} \textbf{\bibinfo{volume}{36}},
  \bibinfo{pages}{2150146} (\bibinfo{year}{2021}), \eprint{2001.00452}.

\bibitem[{\citenamefont{Mukherjee et~al.}(2017)\citenamefont{Mukherjee, Borah,
  and Das}}]{Mukherjee:2017pzq}
\bibinfo{author}{\bibfnamefont{A.}~\bibnamefont{Mukherjee}},
  \bibinfo{author}{\bibfnamefont{D.}~\bibnamefont{Borah}}, \bibnamefont{and}
  \bibinfo{author}{\bibfnamefont{M.~K.} \bibnamefont{Das}},
  \bibinfo{journal}{Phys. Rev. D} \textbf{\bibinfo{volume}{96}},
  \bibinfo{pages}{015014} (\bibinfo{year}{2017}), \eprint{1703.06750}.

\bibitem[{\citenamefont{Heavens and Sellentin}(2018)}]{Heavens:2018adv}
\bibinfo{author}{\bibfnamefont{A.~F.} \bibnamefont{Heavens}} \bibnamefont{and}
  \bibinfo{author}{\bibfnamefont{E.}~\bibnamefont{Sellentin}},
  \bibinfo{journal}{JCAP} \textbf{\bibinfo{volume}{04}}, \bibinfo{pages}{047}
  (\bibinfo{year}{2018}), \eprint{1802.09450}.

\bibitem[{\citenamefont{Jimenez et~al.}(2022)\citenamefont{Jimenez, Pena-Garay,
  Short, Simpson, and Verde}}]{Jimenez:2022dkn}
\bibinfo{author}{\bibfnamefont{R.}~\bibnamefont{Jimenez}},
  \bibinfo{author}{\bibfnamefont{C.}~\bibnamefont{Pena-Garay}},
  \bibinfo{author}{\bibfnamefont{K.}~\bibnamefont{Short}},
  \bibinfo{author}{\bibfnamefont{F.}~\bibnamefont{Simpson}}, \bibnamefont{and}
  \bibinfo{author}{\bibfnamefont{L.}~\bibnamefont{Verde}},
  \bibinfo{journal}{JCAP} \textbf{\bibinfo{volume}{09}}, \bibinfo{pages}{006}
  (\bibinfo{year}{2022}), \eprint{2203.14247}.

\bibitem[{\citenamefont{Gariazzo et~al.}(2022)}]{Gariazzo:2022ahe}
\bibinfo{author}{\bibfnamefont{S.}~\bibnamefont{Gariazzo}}
  \bibnamefont{et~al.}, \bibinfo{journal}{JCAP} \textbf{\bibinfo{volume}{10}},
  \bibinfo{pages}{010} (\bibinfo{year}{2022}), \eprint{2205.02195}.

\bibitem[{\citenamefont{Fraser et~al.}(2014)\citenamefont{Fraser, Ma, and
  Popov}}]{Fraser:2014yha}
\bibinfo{author}{\bibfnamefont{S.}~\bibnamefont{Fraser}},
  \bibinfo{author}{\bibfnamefont{E.}~\bibnamefont{Ma}}, \bibnamefont{and}
  \bibinfo{author}{\bibfnamefont{O.}~\bibnamefont{Popov}},
  \bibinfo{journal}{Phys. Lett. B} \textbf{\bibinfo{volume}{737}},
  \bibinfo{pages}{280} (\bibinfo{year}{2014}), \eprint{1408.4785}.

\bibitem[{\citenamefont{Mandal et~al.}(2021{\natexlab{a}})\citenamefont{Mandal,
  Srivastava, and Valle}}]{Mandal:2021yph}
\bibinfo{author}{\bibfnamefont{S.}~\bibnamefont{Mandal}},
  \bibinfo{author}{\bibfnamefont{R.}~\bibnamefont{Srivastava}},
  \bibnamefont{and} \bibinfo{author}{\bibfnamefont{J.~W.~F.}
  \bibnamefont{Valle}}, \bibinfo{journal}{Phys. Lett. B}
  \textbf{\bibinfo{volume}{819}}, \bibinfo{pages}{136458}
  (\bibinfo{year}{2021}{\natexlab{a}}), \eprint{2104.13401}.

\bibitem[{\citenamefont{Mandal et~al.}(2021{\natexlab{b}})\citenamefont{Mandal,
  Rojas, Srivastava, and Valle}}]{Mandal:2019oth}
\bibinfo{author}{\bibfnamefont{S.}~\bibnamefont{Mandal}},
  \bibinfo{author}{\bibfnamefont{N.}~\bibnamefont{Rojas}},
  \bibinfo{author}{\bibfnamefont{R.}~\bibnamefont{Srivastava}},
  \bibnamefont{and} \bibinfo{author}{\bibfnamefont{J.~W.~F.}
  \bibnamefont{Valle}}, \bibinfo{journal}{Phys. Lett. B}
  \textbf{\bibinfo{volume}{821}}, \bibinfo{pages}{136609}
  (\bibinfo{year}{2021}{\natexlab{b}}), \eprint{1907.07728}.

\bibitem[{\citenamefont{Feruglio}(2019)}]{Feruglio:2017spp}
\bibinfo{author}{\bibfnamefont{F.}~\bibnamefont{Feruglio}},
  \emph{\bibinfo{title}{{Are neutrino masses modular forms?}}}
  (\bibinfo{year}{2019}), pp. \bibinfo{pages}{227--266}, \eprint{1706.08749}.

\bibitem[{\citenamefont{King and King}(2020)}]{King:2020qaj}
\bibinfo{author}{\bibfnamefont{S.~J.~D.} \bibnamefont{King}} \bibnamefont{and}
  \bibinfo{author}{\bibfnamefont{S.~F.} \bibnamefont{King}},
  \bibinfo{journal}{JHEP} \textbf{\bibinfo{volume}{09}}, \bibinfo{pages}{043}
  (\bibinfo{year}{2020}), \eprint{2002.00969}.

\bibitem[{\citenamefont{Kobayashi et~al.}(2018)\citenamefont{Kobayashi, Tanaka,
  and Tatsuishi}}]{Kobayashi:2018vbk}
\bibinfo{author}{\bibfnamefont{T.}~\bibnamefont{Kobayashi}},
  \bibinfo{author}{\bibfnamefont{K.}~\bibnamefont{Tanaka}}, \bibnamefont{and}
  \bibinfo{author}{\bibfnamefont{T.~H.} \bibnamefont{Tatsuishi}},
  \bibinfo{journal}{Phys. Rev. D} \textbf{\bibinfo{volume}{98}},
  \bibinfo{pages}{016004} (\bibinfo{year}{2018}), \eprint{1803.10391}.

\bibitem[{\citenamefont{Okada and Orikasa}(2019)}]{Okada:2019xqk}
\bibinfo{author}{\bibfnamefont{H.}~\bibnamefont{Okada}} \bibnamefont{and}
  \bibinfo{author}{\bibfnamefont{Y.}~\bibnamefont{Orikasa}},
  \bibinfo{journal}{Phys. Rev. D} \textbf{\bibinfo{volume}{100}},
  \bibinfo{pages}{115037} (\bibinfo{year}{2019}), \eprint{1907.04716}.

\bibitem[{\citenamefont{Meloni and Parriciatu}(2023)}]{Meloni:2023aru}
\bibinfo{author}{\bibfnamefont{D.}~\bibnamefont{Meloni}} \bibnamefont{and}
  \bibinfo{author}{\bibfnamefont{M.}~\bibnamefont{Parriciatu}},
  \bibinfo{journal}{JHEP} \textbf{\bibinfo{volume}{09}}, \bibinfo{pages}{043}
  (\bibinfo{year}{2023}), \eprint{2306.09028}.

\bibitem[{\citenamefont{Kobayashi et~al.}(2019)\citenamefont{Kobayashi,
  Shimizu, Takagi, Tanimoto, and Tatsuishi}}]{Kobayashi:2019xvz}
\bibinfo{author}{\bibfnamefont{T.}~\bibnamefont{Kobayashi}},
  \bibinfo{author}{\bibfnamefont{Y.}~\bibnamefont{Shimizu}},
  \bibinfo{author}{\bibfnamefont{K.}~\bibnamefont{Takagi}},
  \bibinfo{author}{\bibfnamefont{M.}~\bibnamefont{Tanimoto}}, \bibnamefont{and}
  \bibinfo{author}{\bibfnamefont{T.~H.} \bibnamefont{Tatsuishi}},
  \bibinfo{journal}{Phys. Rev. D} \textbf{\bibinfo{volume}{100}},
  \bibinfo{pages}{115045} (\bibinfo{year}{2019}), \bibinfo{note}{[Erratum:
  Phys.Rev.D 101, 039904 (2020)]}, \eprint{1909.05139}.

\bibitem[{\citenamefont{Wang and Zhou}(2020)}]{Wang:2019ovr}
\bibinfo{author}{\bibfnamefont{X.}~\bibnamefont{Wang}} \bibnamefont{and}
  \bibinfo{author}{\bibfnamefont{S.}~\bibnamefont{Zhou}},
  \bibinfo{journal}{JHEP} \textbf{\bibinfo{volume}{05}}, \bibinfo{pages}{017}
  (\bibinfo{year}{2020}), \eprint{1910.09473}.

\bibitem[{\citenamefont{Penedo and Petcov}(2019)}]{Penedo:2018nmg}
\bibinfo{author}{\bibfnamefont{J.~T.} \bibnamefont{Penedo}} \bibnamefont{and}
  \bibinfo{author}{\bibfnamefont{S.~T.} \bibnamefont{Petcov}},
  \bibinfo{journal}{Nucl. Phys. B} \textbf{\bibinfo{volume}{939}},
  \bibinfo{pages}{292} (\bibinfo{year}{2019}), \eprint{1806.11040}.

\bibitem[{\citenamefont{Zhang and Zhou}(2021)}]{Zhang:2021olk}
\bibinfo{author}{\bibfnamefont{X.}~\bibnamefont{Zhang}} \bibnamefont{and}
  \bibinfo{author}{\bibfnamefont{S.}~\bibnamefont{Zhou}},
  \bibinfo{journal}{JCAP} \textbf{\bibinfo{volume}{09}}, \bibinfo{pages}{043}
  (\bibinfo{year}{2021}), \eprint{2106.03433}.

\bibitem[{\citenamefont{Kashav and Verma}(2021)}]{Kashav:2021zir}
\bibinfo{author}{\bibfnamefont{M.}~\bibnamefont{Kashav}} \bibnamefont{and}
  \bibinfo{author}{\bibfnamefont{S.}~\bibnamefont{Verma}},
  \bibinfo{journal}{JHEP} \textbf{\bibinfo{volume}{09}}, \bibinfo{pages}{100}
  (\bibinfo{year}{2021}), \eprint{2103.07207}.

\bibitem[{\citenamefont{Kashav and Verma}(2023)}]{Kashav:2022kpk}
\bibinfo{author}{\bibfnamefont{M.}~\bibnamefont{Kashav}} \bibnamefont{and}
  \bibinfo{author}{\bibfnamefont{S.}~\bibnamefont{Verma}},
  \bibinfo{journal}{JCAP} \textbf{\bibinfo{volume}{03}}, \bibinfo{pages}{010}
  (\bibinfo{year}{2023}), \eprint{2205.06545}.

\bibitem[{\citenamefont{Singh et~al.}(2024)\citenamefont{Singh, Kashav, and
  Verma}}]{Singh:2024imk}
\bibinfo{author}{\bibfnamefont{L.}~\bibnamefont{Singh}},
  \bibinfo{author}{\bibfnamefont{M.}~\bibnamefont{Kashav}}, \bibnamefont{and}
  \bibinfo{author}{\bibfnamefont{S.}~\bibnamefont{Verma}},
  \bibinfo{journal}{Nucl. Phys. B} \textbf{\bibinfo{volume}{1007}},
  \bibinfo{pages}{116666} (\bibinfo{year}{2024}), \eprint{2405.07165}.

\bibitem[{\citenamefont{Mishra et~al.}(2023)\citenamefont{Mishra, Behera, and
  Mohanta}}]{Mishra:2023cjc}
\bibinfo{author}{\bibfnamefont{P.}~\bibnamefont{Mishra}},
  \bibinfo{author}{\bibfnamefont{M.~K.} \bibnamefont{Behera}},
  \bibnamefont{and} \bibinfo{author}{\bibfnamefont{R.}~\bibnamefont{Mohanta}},
  \bibinfo{journal}{Phys. Rev. D} \textbf{\bibinfo{volume}{107}},
  \bibinfo{pages}{115004} (\bibinfo{year}{2023}), \eprint{2302.00494}.

\bibitem[{\citenamefont{Nomura and Okada}(2019)}]{Nomura:2019jxj}
\bibinfo{author}{\bibfnamefont{T.}~\bibnamefont{Nomura}} \bibnamefont{and}
  \bibinfo{author}{\bibfnamefont{H.}~\bibnamefont{Okada}},
  \bibinfo{journal}{Phys. Lett. B} \textbf{\bibinfo{volume}{797}},
  \bibinfo{pages}{134799} (\bibinfo{year}{2019}), \eprint{1904.03937}.

\bibitem[{\citenamefont{Nomura et~al.}(2021)\citenamefont{Nomura, Okada, and
  Patra}}]{Nomura:2019xsb}
\bibinfo{author}{\bibfnamefont{T.}~\bibnamefont{Nomura}},
  \bibinfo{author}{\bibfnamefont{H.}~\bibnamefont{Okada}}, \bibnamefont{and}
  \bibinfo{author}{\bibfnamefont{S.}~\bibnamefont{Patra}},
  \bibinfo{journal}{Nucl. Phys. B} \textbf{\bibinfo{volume}{967}},
  \bibinfo{pages}{115395} (\bibinfo{year}{2021}), \eprint{1912.00379}.

\bibitem[{\citenamefont{Gogoi et~al.}(2024)\citenamefont{Gogoi, Sarma, and
  Das}}]{Gogoi:2023jzl}
\bibinfo{author}{\bibfnamefont{J.}~\bibnamefont{Gogoi}},
  \bibinfo{author}{\bibfnamefont{L.}~\bibnamefont{Sarma}}, \bibnamefont{and}
  \bibinfo{author}{\bibfnamefont{M.~K.} \bibnamefont{Das}},
  \bibinfo{journal}{Eur. Phys. J. C} \textbf{\bibinfo{volume}{84}},
  \bibinfo{pages}{689} (\bibinfo{year}{2024}), \eprint{2311.09883}.

\bibitem[{\citenamefont{Behera et~al.}(2022{\natexlab{a}})\citenamefont{Behera,
  Singirala, Mishra, and Mohanta}}]{Behera:2020lpd}
\bibinfo{author}{\bibfnamefont{M.~K.} \bibnamefont{Behera}},
  \bibinfo{author}{\bibfnamefont{S.}~\bibnamefont{Singirala}},
  \bibinfo{author}{\bibfnamefont{S.}~\bibnamefont{Mishra}}, \bibnamefont{and}
  \bibinfo{author}{\bibfnamefont{R.}~\bibnamefont{Mohanta}},
  \bibinfo{journal}{J. Phys. G} \textbf{\bibinfo{volume}{49}},
  \bibinfo{pages}{035002} (\bibinfo{year}{2022}{\natexlab{a}}),
  \eprint{2009.01806}.

\bibitem[{\citenamefont{Abbas}(2021)}]{Abbas:2020qzc}
\bibinfo{author}{\bibfnamefont{M.}~\bibnamefont{Abbas}},
  \bibinfo{journal}{Phys. Rev. D} \textbf{\bibinfo{volume}{103}},
  \bibinfo{pages}{056016} (\bibinfo{year}{2021}), \eprint{2002.01929}.

\bibitem[{\citenamefont{Altarelli and Feruglio}(2006)}]{Altarelli:2005yx}
\bibinfo{author}{\bibfnamefont{G.}~\bibnamefont{Altarelli}} \bibnamefont{and}
  \bibinfo{author}{\bibfnamefont{F.}~\bibnamefont{Feruglio}},
  \bibinfo{journal}{Nucl. Phys. B} \textbf{\bibinfo{volume}{741}},
  \bibinfo{pages}{215} (\bibinfo{year}{2006}), \eprint{hep-ph/0512103}.

\bibitem[{\citenamefont{Das et~al.}(2019)\citenamefont{Das, Mukherjee, and
  Das}}]{Das:2018qyt}
\bibinfo{author}{\bibfnamefont{P.}~\bibnamefont{Das}},
  \bibinfo{author}{\bibfnamefont{A.}~\bibnamefont{Mukherjee}},
  \bibnamefont{and} \bibinfo{author}{\bibfnamefont{M.~K.} \bibnamefont{Das}},
  \bibinfo{journal}{Nucl. Phys. B} \textbf{\bibinfo{volume}{941}},
  \bibinfo{pages}{755} (\bibinfo{year}{2019}), \eprint{1805.09231}.

\bibitem[{\citenamefont{Novichkov et~al.}(2019)\citenamefont{Novichkov, Penedo,
  Petcov, and Titov}}]{Novichkov:2018nkm}
\bibinfo{author}{\bibfnamefont{P.~P.} \bibnamefont{Novichkov}},
  \bibinfo{author}{\bibfnamefont{J.~T.} \bibnamefont{Penedo}},
  \bibinfo{author}{\bibfnamefont{S.~T.} \bibnamefont{Petcov}},
  \bibnamefont{and} \bibinfo{author}{\bibfnamefont{A.~V.} \bibnamefont{Titov}},
  \bibinfo{journal}{JHEP} \textbf{\bibinfo{volume}{04}}, \bibinfo{pages}{174}
  (\bibinfo{year}{2019}), \eprint{1812.02158}.

\bibitem[{\citenamefont{Ding et~al.}(2019)\citenamefont{Ding, King, and
  Liu}}]{Ding:2019xna}
\bibinfo{author}{\bibfnamefont{G.-J.} \bibnamefont{Ding}},
  \bibinfo{author}{\bibfnamefont{S.~F.} \bibnamefont{King}}, \bibnamefont{and}
  \bibinfo{author}{\bibfnamefont{X.-G.} \bibnamefont{Liu}},
  \bibinfo{journal}{Phys. Rev. D} \textbf{\bibinfo{volume}{100}},
  \bibinfo{pages}{115005} (\bibinfo{year}{2019}), \eprint{1903.12588}.

\bibitem[{\citenamefont{Behera and
  Mohanta}(2022{\natexlab{a}})}]{Behera:2021eut}
\bibinfo{author}{\bibfnamefont{M.~K.} \bibnamefont{Behera}} \bibnamefont{and}
  \bibinfo{author}{\bibfnamefont{R.}~\bibnamefont{Mohanta}},
  \bibinfo{journal}{J. Phys. G} \textbf{\bibinfo{volume}{49}},
  \bibinfo{pages}{045001} (\bibinfo{year}{2022}{\natexlab{a}}),
  \eprint{2108.01059}.

\bibitem[{\citenamefont{Behera and
  Mohanta}(2022{\natexlab{b}})}]{Behera:2022wco}
\bibinfo{author}{\bibfnamefont{M.~K.} \bibnamefont{Behera}} \bibnamefont{and}
  \bibinfo{author}{\bibfnamefont{R.}~\bibnamefont{Mohanta}},
  \bibinfo{journal}{Front. in Phys.} \textbf{\bibinfo{volume}{10}},
  \bibinfo{pages}{854595} (\bibinfo{year}{2022}{\natexlab{b}}),
  \eprint{2201.10429}.

\bibitem[{\citenamefont{Kumar et~al.}(2024{\natexlab{a}})\citenamefont{Kumar,
  Nath, and Srivastava}}]{Kumar:2024zfb}
\bibinfo{author}{\bibfnamefont{R.}~\bibnamefont{Kumar}},
  \bibinfo{author}{\bibfnamefont{N.}~\bibnamefont{Nath}}, \bibnamefont{and}
  \bibinfo{author}{\bibfnamefont{R.}~\bibnamefont{Srivastava}}
  (\bibinfo{year}{2024}{\natexlab{a}}), \eprint{2406.00188}.

\bibitem[{\citenamefont{Kumar et~al.}(2024{\natexlab{b}})\citenamefont{Kumar,
  Mishra, Behera, Mohanta, and Srivastava}}]{Kumar:2023moh}
\bibinfo{author}{\bibfnamefont{R.}~\bibnamefont{Kumar}},
  \bibinfo{author}{\bibfnamefont{P.}~\bibnamefont{Mishra}},
  \bibinfo{author}{\bibfnamefont{M.~K.} \bibnamefont{Behera}},
  \bibinfo{author}{\bibfnamefont{R.}~\bibnamefont{Mohanta}}, \bibnamefont{and}
  \bibinfo{author}{\bibfnamefont{R.}~\bibnamefont{Srivastava}},
  \bibinfo{journal}{Phys. Lett. B} \textbf{\bibinfo{volume}{853}},
  \bibinfo{pages}{138635} (\bibinfo{year}{2024}{\natexlab{b}}),
  \eprint{2310.02363}.

\bibitem[{\citenamefont{Behera et~al.}(2022{\natexlab{b}})\citenamefont{Behera,
  Mishra, Singirala, and Mohanta}}]{Behera:2020sfe}
\bibinfo{author}{\bibfnamefont{M.~K.} \bibnamefont{Behera}},
  \bibinfo{author}{\bibfnamefont{S.}~\bibnamefont{Mishra}},
  \bibinfo{author}{\bibfnamefont{S.}~\bibnamefont{Singirala}},
  \bibnamefont{and} \bibinfo{author}{\bibfnamefont{R.}~\bibnamefont{Mohanta}},
  \bibinfo{journal}{Phys. Dark Univ.} \textbf{\bibinfo{volume}{36}},
  \bibinfo{pages}{101027} (\bibinfo{year}{2022}{\natexlab{b}}),
  \eprint{2007.00545}.

\bibitem[{\citenamefont{Centelles~Chuli\'a
  et~al.}(2024)\citenamefont{Centelles~Chuli\'a, Kumar, Popov, and
  Srivastava}}]{CentellesChulia:2023osj}
\bibinfo{author}{\bibfnamefont{S.}~\bibnamefont{Centelles~Chuli\'a}},
  \bibinfo{author}{\bibfnamefont{R.}~\bibnamefont{Kumar}},
  \bibinfo{author}{\bibfnamefont{O.}~\bibnamefont{Popov}}, \bibnamefont{and}
  \bibinfo{author}{\bibfnamefont{R.}~\bibnamefont{Srivastava}},
  \bibinfo{journal}{Phys. Rev. D} \textbf{\bibinfo{volume}{109}},
  \bibinfo{pages}{035016} (\bibinfo{year}{2024}), \eprint{2308.08981}.

\bibitem[{\citenamefont{Mishra et~al.}(2022)\citenamefont{Mishra, Behera,
  Panda, and Mohanta}}]{Mishra:2022egy}
\bibinfo{author}{\bibfnamefont{P.}~\bibnamefont{Mishra}},
  \bibinfo{author}{\bibfnamefont{M.~K.} \bibnamefont{Behera}},
  \bibinfo{author}{\bibfnamefont{P.}~\bibnamefont{Panda}}, \bibnamefont{and}
  \bibinfo{author}{\bibfnamefont{R.}~\bibnamefont{Mohanta}},
  \bibinfo{journal}{Eur. Phys. J. C} \textbf{\bibinfo{volume}{82}},
  \bibinfo{pages}{1115} (\bibinfo{year}{2022}), \eprint{2204.08338}.

\bibitem[{\citenamefont{Esteban et~al.}(2020)\citenamefont{Esteban,
  Gonzalez-Garcia, Maltoni, Schwetz, and Zhou}}]{Esteban:2020cvm}
\bibinfo{author}{\bibfnamefont{I.}~\bibnamefont{Esteban}},
  \bibinfo{author}{\bibfnamefont{M.~C.} \bibnamefont{Gonzalez-Garcia}},
  \bibinfo{author}{\bibfnamefont{M.}~\bibnamefont{Maltoni}},
  \bibinfo{author}{\bibfnamefont{T.}~\bibnamefont{Schwetz}}, \bibnamefont{and}
  \bibinfo{author}{\bibfnamefont{A.}~\bibnamefont{Zhou}},
  \bibinfo{journal}{JHEP} \textbf{\bibinfo{volume}{09}}, \bibinfo{pages}{178}
  (\bibinfo{year}{2020}), \eprint{2007.14792}.

\bibitem[{\citenamefont{Alloul et~al.}(2014)\citenamefont{Alloul, Christensen,
  Degrande, Duhr, and Fuks}}]{Alloul:2013bka}
\bibinfo{author}{\bibfnamefont{A.}~\bibnamefont{Alloul}},
  \bibinfo{author}{\bibfnamefont{N.~D.} \bibnamefont{Christensen}},
  \bibinfo{author}{\bibfnamefont{C.}~\bibnamefont{Degrande}},
  \bibinfo{author}{\bibfnamefont{C.}~\bibnamefont{Duhr}}, \bibnamefont{and}
  \bibinfo{author}{\bibfnamefont{B.}~\bibnamefont{Fuks}},
  \bibinfo{journal}{Comput. Phys. Commun.} \textbf{\bibinfo{volume}{185}},
  \bibinfo{pages}{2250} (\bibinfo{year}{2014}), \eprint{1310.1921}.

\bibitem[{\citenamefont{Alguero et~al.}(2024)\citenamefont{Alguero, Belanger,
  Boudjema, Chakraborti, Goudelis, Kraml, Mjallal, and
  Pukhov}}]{Alguero:2023zol}
\bibinfo{author}{\bibfnamefont{G.}~\bibnamefont{Alguero}},
  \bibinfo{author}{\bibfnamefont{G.}~\bibnamefont{Belanger}},
  \bibinfo{author}{\bibfnamefont{F.}~\bibnamefont{Boudjema}},
  \bibinfo{author}{\bibfnamefont{S.}~\bibnamefont{Chakraborti}},
  \bibinfo{author}{\bibfnamefont{A.}~\bibnamefont{Goudelis}},
  \bibinfo{author}{\bibfnamefont{S.}~\bibnamefont{Kraml}},
  \bibinfo{author}{\bibfnamefont{A.}~\bibnamefont{Mjallal}}, \bibnamefont{and}
  \bibinfo{author}{\bibfnamefont{A.}~\bibnamefont{Pukhov}},
  \bibinfo{journal}{Comput. Phys. Commun.} \textbf{\bibinfo{volume}{299}},
  \bibinfo{pages}{109133} (\bibinfo{year}{2024}), \eprint{2312.14894}.

\bibitem[{\citenamefont{Chun et~al.}(2023)\citenamefont{Chun, Roy, Mandal, and
  Mitra}}]{Chun:2023vbh}
\bibinfo{author}{\bibfnamefont{E.~J.} \bibnamefont{Chun}},
  \bibinfo{author}{\bibfnamefont{A.}~\bibnamefont{Roy}},
  \bibinfo{author}{\bibfnamefont{S.}~\bibnamefont{Mandal}}, \bibnamefont{and}
  \bibinfo{author}{\bibfnamefont{M.}~\bibnamefont{Mitra}},
  \bibinfo{journal}{JHEP} \textbf{\bibinfo{volume}{08}}, \bibinfo{pages}{130}
  (\bibinfo{year}{2023}), \eprint{2303.02681}.

\bibitem[{\citenamefont{Karan et~al.}(2023)\citenamefont{Karan, Sadhukhan, and
  Valle}}]{Karan:2023adm}
\bibinfo{author}{\bibfnamefont{A.}~\bibnamefont{Karan}},
  \bibinfo{author}{\bibfnamefont{S.}~\bibnamefont{Sadhukhan}},
  \bibnamefont{and} \bibinfo{author}{\bibfnamefont{J.~W.~F.}
  \bibnamefont{Valle}}, \bibinfo{journal}{JHEP} \textbf{\bibinfo{volume}{12}},
  \bibinfo{pages}{185} (\bibinfo{year}{2023}), \eprint{2308.09135}.

\bibitem[{\citenamefont{Borah et~al.}(2024{\natexlab{a}})\citenamefont{Borah,
  Das, Karmakar, and Mahapatra}}]{Borah:2024gql}
\bibinfo{author}{\bibfnamefont{D.}~\bibnamefont{Borah}},
  \bibinfo{author}{\bibfnamefont{P.}~\bibnamefont{Das}},
  \bibinfo{author}{\bibfnamefont{B.}~\bibnamefont{Karmakar}}, \bibnamefont{and}
  \bibinfo{author}{\bibfnamefont{S.}~\bibnamefont{Mahapatra}}
  (\bibinfo{year}{2024}{\natexlab{a}}), \eprint{2406.17861}.

\bibitem[{\citenamefont{Borah et~al.}(2024{\natexlab{b}})\citenamefont{Borah,
  Das, and Nanda}}]{Borah:2022enh}
\bibinfo{author}{\bibfnamefont{D.}~\bibnamefont{Borah}},
  \bibinfo{author}{\bibfnamefont{P.}~\bibnamefont{Das}}, \bibnamefont{and}
  \bibinfo{author}{\bibfnamefont{D.}~\bibnamefont{Nanda}},
  \bibinfo{journal}{Eur. Phys. J. C} \textbf{\bibinfo{volume}{84}},
  \bibinfo{pages}{140} (\bibinfo{year}{2024}{\natexlab{b}}),
  \eprint{2211.13168}.

\bibitem[{\citenamefont{Ganguly et~al.}(2024)\citenamefont{Ganguly, Gluza,
  Karmakar, and Mahapatra}}]{Ganguly:2023jml}
\bibinfo{author}{\bibfnamefont{J.}~\bibnamefont{Ganguly}},
  \bibinfo{author}{\bibfnamefont{J.}~\bibnamefont{Gluza}},
  \bibinfo{author}{\bibfnamefont{B.}~\bibnamefont{Karmakar}}, \bibnamefont{and}
  \bibinfo{author}{\bibfnamefont{S.}~\bibnamefont{Mahapatra}},
  \bibinfo{journal}{Phys. Rev. D} \textbf{\bibinfo{volume}{110}},
  \bibinfo{pages}{035012} (\bibinfo{year}{2024}), \eprint{2311.15997}.

\bibitem[{\citenamefont{Borah et~al.}(2024{\natexlab{c}})\citenamefont{Borah,
  Mahapatra, Paul, and Sahu}}]{Borah:2023hqw}
\bibinfo{author}{\bibfnamefont{D.}~\bibnamefont{Borah}},
  \bibinfo{author}{\bibfnamefont{S.}~\bibnamefont{Mahapatra}},
  \bibinfo{author}{\bibfnamefont{P.~K.} \bibnamefont{Paul}}, \bibnamefont{and}
  \bibinfo{author}{\bibfnamefont{N.}~\bibnamefont{Sahu}},
  \bibinfo{journal}{Phys. Rev. D} \textbf{\bibinfo{volume}{109}},
  \bibinfo{pages}{055021} (\bibinfo{year}{2024}{\natexlab{c}}),
  \eprint{2310.11953}.

\bibitem[{\citenamefont{Abe et~al.}(2023)}]{KamLAND-Zen:2022tow}
\bibinfo{author}{\bibfnamefont{S.}~\bibnamefont{Abe}} \bibnamefont{et~al.}
  (\bibinfo{collaboration}{KamLAND-Zen}), \bibinfo{journal}{Phys. Rev. Lett.}
  \textbf{\bibinfo{volume}{130}}, \bibinfo{pages}{051801}
  (\bibinfo{year}{2023}), \eprint{2203.02139}.

\bibitem[{\citenamefont{Andringa et~al.}(2016)}]{SNO:2015wyx}
\bibinfo{author}{\bibfnamefont{S.}~\bibnamefont{Andringa}} \bibnamefont{et~al.}
  (\bibinfo{collaboration}{SNO+}), \bibinfo{journal}{Adv. High Energy Phys.}
  \textbf{\bibinfo{volume}{2016}}, \bibinfo{pages}{6194250}
  (\bibinfo{year}{2016}), \eprint{1508.05759}.

\bibitem[{\citenamefont{Abgrall et~al.}(2021)}]{LEGEND:2021bnm}
\bibinfo{author}{\bibfnamefont{N.}~\bibnamefont{Abgrall}} \bibnamefont{et~al.}
  (\bibinfo{collaboration}{LEGEND}) (\bibinfo{year}{2021}),
  \eprint{2107.11462}.

\bibitem[{\citenamefont{Albert et~al.}(2018)}]{nEXO:2017nam}
\bibinfo{author}{\bibfnamefont{J.~B.} \bibnamefont{Albert}}
  \bibnamefont{et~al.} (\bibinfo{collaboration}{nEXO}), \bibinfo{journal}{Phys.
  Rev. C} \textbf{\bibinfo{volume}{97}}, \bibinfo{pages}{065503}
  (\bibinfo{year}{2018}), \eprint{1710.05075}.

\bibitem[{\citenamefont{Calibbi and Signorelli}(2018)}]{Calibbi:2017uvl}
\bibinfo{author}{\bibfnamefont{L.}~\bibnamefont{Calibbi}} \bibnamefont{and}
  \bibinfo{author}{\bibfnamefont{G.}~\bibnamefont{Signorelli}},
  \bibinfo{journal}{Riv. Nuovo Cim.} \textbf{\bibinfo{volume}{41}},
  \bibinfo{pages}{71} (\bibinfo{year}{2018}), \eprint{1709.00294}.

\bibitem[{\citenamefont{Ardu and Pezzullo}(2022)}]{Ardu:2022sbt}
\bibinfo{author}{\bibfnamefont{M.}~\bibnamefont{Ardu}} \bibnamefont{and}
  \bibinfo{author}{\bibfnamefont{G.}~\bibnamefont{Pezzullo}},
  \bibinfo{journal}{Universe} \textbf{\bibinfo{volume}{8}},
  \bibinfo{pages}{299} (\bibinfo{year}{2022}), \eprint{2204.08220}.

\bibitem[{\citenamefont{Chakraborty et~al.}(2022)\citenamefont{Chakraborty,
  Roy, and Srivastava}}]{Chakraborty:2021azg}
\bibinfo{author}{\bibfnamefont{I.}~\bibnamefont{Chakraborty}},
  \bibinfo{author}{\bibfnamefont{H.}~\bibnamefont{Roy}}, \bibnamefont{and}
  \bibinfo{author}{\bibfnamefont{T.}~\bibnamefont{Srivastava}},
  \bibinfo{journal}{Nucl. Phys. B} \textbf{\bibinfo{volume}{979}},
  \bibinfo{pages}{115780} (\bibinfo{year}{2022}), \eprint{2106.08232}.

\bibitem[{\citenamefont{Deppisch and Valle}(2005)}]{Deppisch:2004fa}
\bibinfo{author}{\bibfnamefont{F.}~\bibnamefont{Deppisch}} \bibnamefont{and}
  \bibinfo{author}{\bibfnamefont{J.~W.~F.} \bibnamefont{Valle}},
  \bibinfo{journal}{Phys. Rev. D} \textbf{\bibinfo{volume}{72}},
  \bibinfo{pages}{036001} (\bibinfo{year}{2005}), \eprint{hep-ph/0406040}.

\bibitem[{\citenamefont{Forero et~al.}(2011)\citenamefont{Forero, Morisi,
  Tortola, and Valle}}]{Forero:2011pc}
\bibinfo{author}{\bibfnamefont{D.~V.} \bibnamefont{Forero}},
  \bibinfo{author}{\bibfnamefont{S.}~\bibnamefont{Morisi}},
  \bibinfo{author}{\bibfnamefont{M.}~\bibnamefont{Tortola}}, \bibnamefont{and}
  \bibinfo{author}{\bibfnamefont{J.~W.~F.} \bibnamefont{Valle}},
  \bibinfo{journal}{JHEP} \textbf{\bibinfo{volume}{09}}, \bibinfo{pages}{142}
  (\bibinfo{year}{2011}), \eprint{1107.6009}.

\bibitem[{\citenamefont{Chekkal et~al.}(2017)\citenamefont{Chekkal, Ahriche,
  Hammou, and Nasri}}]{Chekkal:2017eka}
\bibinfo{author}{\bibfnamefont{M.}~\bibnamefont{Chekkal}},
  \bibinfo{author}{\bibfnamefont{A.}~\bibnamefont{Ahriche}},
  \bibinfo{author}{\bibfnamefont{A.~B.} \bibnamefont{Hammou}},
  \bibnamefont{and} \bibinfo{author}{\bibfnamefont{S.}~\bibnamefont{Nasri}},
  \bibinfo{journal}{Phys. Rev. D} \textbf{\bibinfo{volume}{95}},
  \bibinfo{pages}{095025} (\bibinfo{year}{2017}), \eprint{1702.04399}.

\bibitem[{\citenamefont{Ilakovac and Pilaftsis}(1995)}]{Ilakovac:1994kj}
\bibinfo{author}{\bibfnamefont{A.}~\bibnamefont{Ilakovac}} \bibnamefont{and}
  \bibinfo{author}{\bibfnamefont{A.}~\bibnamefont{Pilaftsis}},
  \bibinfo{journal}{Nucl. Phys. B} \textbf{\bibinfo{volume}{437}},
  \bibinfo{pages}{491} (\bibinfo{year}{1995}), \eprint{hep-ph/9403398}.

\bibitem[{\citenamefont{Baldini et~al.}(2016)}]{MEG:2016leq}
\bibinfo{author}{\bibfnamefont{A.~M.} \bibnamefont{Baldini}}
  \bibnamefont{et~al.} (\bibinfo{collaboration}{MEG}), \bibinfo{journal}{Eur.
  Phys. J. C} \textbf{\bibinfo{volume}{76}}, \bibinfo{pages}{434}
  (\bibinfo{year}{2016}), \eprint{1605.05081}.

\bibitem[{\citenamefont{Baldini et~al.}(2013)}]{Baldini:2013ke}
\bibinfo{author}{\bibfnamefont{A.~M.} \bibnamefont{Baldini}}
  \bibnamefont{et~al.} (\bibinfo{year}{2013}), \eprint{1301.7225}.

\bibitem[{\citenamefont{Aubert et~al.}(2010)}]{BaBar:2009hkt}
\bibinfo{author}{\bibfnamefont{B.}~\bibnamefont{Aubert}} \bibnamefont{et~al.}
  (\bibinfo{collaboration}{BaBar}), \bibinfo{journal}{Phys. Rev. Lett.}
  \textbf{\bibinfo{volume}{104}}, \bibinfo{pages}{021802}
  (\bibinfo{year}{2010}), \eprint{0908.2381}.

\bibitem[{\citenamefont{Cline et~al.}(2013)\citenamefont{Cline, Kainulainen,
  Scott, and Weniger}}]{Cline:2013gha}
\bibinfo{author}{\bibfnamefont{J.~M.} \bibnamefont{Cline}},
  \bibinfo{author}{\bibfnamefont{K.}~\bibnamefont{Kainulainen}},
  \bibinfo{author}{\bibfnamefont{P.}~\bibnamefont{Scott}}, \bibnamefont{and}
  \bibinfo{author}{\bibfnamefont{C.}~\bibnamefont{Weniger}},
  \bibinfo{journal}{Phys. Rev. D} \textbf{\bibinfo{volume}{88}},
  \bibinfo{pages}{055025} (\bibinfo{year}{2013}), \bibinfo{note}{[Erratum:
  Phys.Rev.D 92, 039906 (2015)]}, \eprint{1306.4710}.

\bibitem[{\citenamefont{Aalbers et~al.}(2023)}]{LZ:2022lsv}
\bibinfo{author}{\bibfnamefont{J.}~\bibnamefont{Aalbers}} \bibnamefont{et~al.}
  (\bibinfo{collaboration}{LZ}), \bibinfo{journal}{Phys. Rev. Lett.}
  \textbf{\bibinfo{volume}{131}}, \bibinfo{pages}{041002}
  (\bibinfo{year}{2023}), \eprint{2207.03764}.

\bibitem[{\citenamefont{Aprile et~al.}(2019)}]{XENON:2019ykp}
\bibinfo{author}{\bibfnamefont{E.}~\bibnamefont{Aprile}} \bibnamefont{et~al.}
  (\bibinfo{collaboration}{XENON}), \bibinfo{journal}{Phys. Rev. D}
  \textbf{\bibinfo{volume}{100}}, \bibinfo{pages}{052014}
  (\bibinfo{year}{2019}), \eprint{1906.04717}.

\bibitem[{\citenamefont{Zhang et~al.}(2019)}]{PandaX:2018wtu}
\bibinfo{author}{\bibfnamefont{H.}~\bibnamefont{Zhang}} \bibnamefont{et~al.}
  (\bibinfo{collaboration}{PandaX}), \bibinfo{journal}{Sci. China Phys. Mech.
  Astron.} \textbf{\bibinfo{volume}{62}}, \bibinfo{pages}{31011}
  (\bibinfo{year}{2019}), \eprint{1806.02229}.

\end{thebibliography}
\end{document}